%% file: Draft_v2a.tex
\documentclass[journal,draftclsnofoot,onecolumn,12pt]{IEEEtran}
\usepackage{amsthm,amssymb,graphicx,multirow,amsmath,color,amsfonts}
\usepackage[update,prepend]{epstopdf}
\usepackage[noadjust]{cite}
\usepackage[latin1]{inputenc}
\usepackage{tikz}
\usetikzlibrary{arrows,calc}		
\usepackage{bbm} 
\usepackage{pdfpages}
\usepackage{tabulary}
\usepackage{multirow}
\usepackage{comment}
\usepackage{graphicx}
\usepackage[font=small,labelfont=bf]{caption}
\usepackage{subcaption}
\usepackage[makeroom]{cancel}
\usepackage{xcolor,etoolbox}
\include{notation}

\allowdisplaybreaks 

\newcommand\cc[1]{\color{black}#1}

\makeatletter 
\pretocmd\@bibitem{\color{black}\csname keycolor#1\endcsname}{}{\fail}
\newcommand\citecolor[1]{\@namedef{keycolor#1}{\color{black}}}
\makeatother
\citecolor{stuber2017principles}
 \citecolor{hager1989updating}
 \citecolor{mengali2013synchronization}  \citecolor{pahlavan2005wireless}
 \citecolor{devineni20multi}
  \citecolor{darsena2019noncoherent}  \citecolor{guruacharya2019optimal}
 
\begin{document}
\graphicspath{{./Figures/}}
\title{Non-coherent Detection and Bit Error Rate for an Ambient Backscatter Link in Time-Selective Fading 
}
\author{
J. Kartheek Devineni and Harpreet S. Dhillon
\thanks{J. K. Devineni and H. S. Dhillon  are with Wireless@VT, Department of ECE, Virginia Tech, Blacksburg, VA (email: kartheekdj@vt.edu and hdhillon@vt.edu).  This work will be presented in part at the IEEE Globecom 2020, Taipei, Taiwan \cite{devineni20multi}. A preliminary study for the special case of dual-antenna receiver was presented in IEEE Globecom 2019 \cite{devineni19noncoherent}. 
\hfill Manuscript updated: \today.} 
}

\maketitle
\vspace{-1.5cm}
\begin{abstract}

This paper focuses on the non-coherent detection in ambient backscatter communication, which is highly appealing for systems where the trade-off between signaling overhead and the actual data transmission is very critical. Modeling the time-selective fading channel as a first-order autoregressive (AR) process, we propose a new receiver architecture based on the direct averaging of the received signal samples for detection, which departs significantly from the energy averaging-based receivers considered in the literature. For the proposed setup, we characterize the exact asymptotic bit error rate (BER) for both single-antenna (SA) and multi-antenna (MA) receivers, and demonstrate the robustness of the new architecture to timing errors. Our results demonstrate that while the direct-link (DL) interference from the ambient power source leads to a BER floor in the SA receiver, the MA receiver can remove this interference by estimating the angle of arrival (AoA) of the DL. The analysis further quantifies the effect of improved angular resolution on the BER as a function of the number of receive antennas. A key intermediate result of our analysis is the derivation of a new concentration result for a general sum sequence that is central to the derivation of the conditional distributions of the received signal.


\end{abstract}


\section{Introduction} \label{sec:intro}

 Ambient backscatter with its technological capability to support battery-free communication has shown remarkable potential in enabling information transfer among energy-constrained devices within the Internet-of-Things (IoT) paradigm \cite{shyam13, shyam16, katti2015}. Given the diverse nature of applications envisioned in the IoT ecosystem, the channel conditions experienced by the IoT devices across different applications could vary significantly \cite{dhillon2015wide}. In the context of this paper, the channel coherence time experienced by these devices could vary by orders of magnitude across applications.  For instance, IoT devices deployed in high mobility scenarios, such as vehicles, road signs, or traffic posts, are expected to experience higher Doppler spread, and hence lower channel coherence time, compared to the IoT devices deployed in relatively static scenarios, such as homes, offices, and public places. While the latter case has implicitly been the focus of most of the prior work on ambient backscatter communication systems, the former is equally, if not more, important but has received much less attention. Most notably, lower coherence time makes it difficult to implement channel estimation and tracking procedure using either training or blind estimation. Because of this, one needs to consider non-coherent detection schemes for such scenarios, which have not yet been investigated in the context of ambient backscatter communications. Motivated by this knowledge gap, this paper focuses on receiver design and comprehensive performance characterization of non-coherent detection-based ambient backscatter system under time-selective fading channels.

\subsection{Related Work} \label{sec:RelW}

As noted above, the existing literature on ambient backscatter is mainly focused on the slow fading channels that assume a block fading model \cite{devineni2019ambient, Wang15, chintha15, chintha16, gao16, gao17, hu2015, chintha16vtc, yang17, el2019noncoherent, tao2018symbol, wang17, zhao2018blind, guo2018noncoherent, zhang2018constellation, guo2018exploiting, darsena2018joint}. Maximum-likelihood (ML) detection under an ambient backscatter setup was first investigated in \cite{Wang15}. The signal detection under non-coherent and semi-coherent setups is analyzed in \cite{chintha15, chintha16, gao16, gao17}. The signal detection at a multiple antenna  receiver is studied in \cite{hu2015} and the statistical-covariance based detection is explored in \cite{chintha16vtc}. While [6]-[12] were based on the Gaussian distribution approximation for the conditional distributions of the average energy of the received signal, the exact BER analysis for the slow fading case was performed in [5]. Interested readers can also refer to [5] for a detailed overview of the backscatter concept. Ambient backscatter communication using orthogonal frequency division multiplexing (OFDM) is investigated in \cite{yang17, el2019noncoherent}. On the same lines, \cite{tao2018symbol, wang17} explored new coding schemes, such as Manchester coding, to improve detection performance. Some of the existing literature that have worked on reducing the affect of the DL interference in ambient backscatter are \cite{shyam14, yang17, el2019noncoherent, duan2019hybrid}. In \cite{shyam14}, a MA prototype is developed which overcomes the affect of DL interference by estimating the channel using the preamble bits of Wi-Fi. In \cite{yang17}, the repetitive pattern of the data in OFDM, due to the use of cyclic prefix, is exploited to cancel the DL interference. Meanwhile, \cite{el2019noncoherent} has designed a non-coherent detector which totally avoids the DL interference by utilizing the null sub-carriers in OFDM. In \cite{duan2019hybrid}, an analog-digital hybrid beamformer receiver, that designs the optimal beamforming vector using the AoA of the DL, is proposed for a deterministic line-of-sight (LOS) channel.  However, these works \cite{shyam14, yang17, el2019noncoherent, duan2019hybrid} consider a block fading channel and none of them have jointly investigated non-coherent detection and time-selective fading which distinguishes our work.

A general requirement of coherent detection is the transmission of pilot/training symbols from transmitter to receiver nodes for the estimation of channel state information (CSI). This will  require some form of cooperation between the primary and backscatter network nodes which might not always be possible. Hence, alternate approaches that avoid the transmission of pilots, such as blind channel estimation techniques, have also been investigated for ambient backscatter \cite{zhao2018blind, guo2018noncoherent, zhang2018constellation, guo2018exploiting, darsena2018joint}. These approaches use different techniques from Bayesian statistics such as expectation-maximization (EM) or space alternating generalized expectation-maximization (SAGE) to iteratively implement the  maximum \emph{a posteriori} probability (MAP) or ML methods to perform the channel estimation from the received signal directly. The performance of these techniques depends on the accuracy and the convergence rate of the blind channel estimation procedures. Therefore, if the convergence rate is slow, these techniques might not be suitable for implementation in a time-selective fading channel. We overcome this drawback by investigating a non-coherent detection technique that only requires estimating large-scale parameters.

\subsection{Contributions} 

To the best of our knowledge, this is the first work that presents a comprehensive analytical treatment of non-coherent detection in ambient backscatter under time-selective fading channels. The time-selective fading channel is modeled using a first-order AR process, and for this setup a binary hypothesis testing problem is formulated to investigate the BER performance of the two following receivers: 1) single-antenna (SA) receiver, and 2) multi-antenna (MA) receiver. 

\paragraph*{New Receiver Architecture}

The receiver architecture used in the prior studies of ambient backscatter requires the computation of the test statistic (TS) based on the average energy of the received signal samples. In our work, we consider a different receiver architecture based on the direct averaging of the received signal samples which requires lesser number of operations, thereby reducing the complexity. Besides, it is more tractable compared to the conventional architecture as derivations for the optimal detection strategies in that detector are not easy \cite{devineni2019ambient}. In addition, due to the exclusive linear operations in the new architecture, the receiver is shown to be resilient to synchronization and timing errors. By deriving BER for the non-coherent setup, we concretely demonstrate that while the new architecture is inadequate for an SA receiver, it has good BER performance when used in conjunction with a MA receiver, which is attributed to the elimination of the strong interference generated by the DL from the power source. The novelty of the MA receiver designed here lies in its ability to exploit the fact that the time-scale over which AoA varies is much larger than the time-scale over which the overall channel gain varies and use it for tracking the AoA of the DL. As implied already, the new receiver architecture also results in tractable conditional distributions, which facilitates the derivation of the optimal detection strategy and the evaluation of the exact BER.

\paragraph*{Asymptotic Growth Rate of a Generalized Sum Sequence}

In the process of deriving conditional distributions, we come across a sum sequence with correlation across  samples. We investigate the asymptotic growth rate of this sum sequence and use it to derive a new concentration result for another specific sequence of interest. This contribution is central to the evaluation of the exact asymptotic conditional distributions and  to the subsequent BER analysis.


\paragraph*{Insights}

 Our analysis has shown that the SA receiver quickly reaches a BER floor due to the strong interference resulting from the DL of the ambient power source. The performance is shown to improve drastically after canceling this interference, which is achieved by tracking the AoA of the DL using the MA receiver. Further, with multiple antennas it is possible to achieve antenna gain, including an additional angular resolution when the the number of receive antennas are increased beyond two. This improvement in angular resolution plays an important role in applications where the AoAs of the DL and backscatter link (BL) are similar. The BER with the new receiver architecture is shown to be independent of the signal sample-size of the averaging operation for some cases, such as zero expected value of the ambient data sequence and/or uncorrelated time-domain fading. For the more general case of correlated fading, the BER is observed to improve with increasing time-domain correlation of the fading. Due to the diminishing returns in the improvement of BER with increasing sample size, the BER initially decreases and then reaches an asymptotic value. In addition, the first-order AR process is shown to be a good approximation of the reference models available for the time-selective fading channels by comparing their BER performance under different scenarios. 


\section{System Model} \label{sec:SysMod}


\subsection{System Setup and Channel Model} \label{sec:ChMod}
 
\begin{figure}
    \centering
    \begin{minipage}[b]{0.51\columnwidth}
        \centering
        \includegraphics [width=\linewidth]{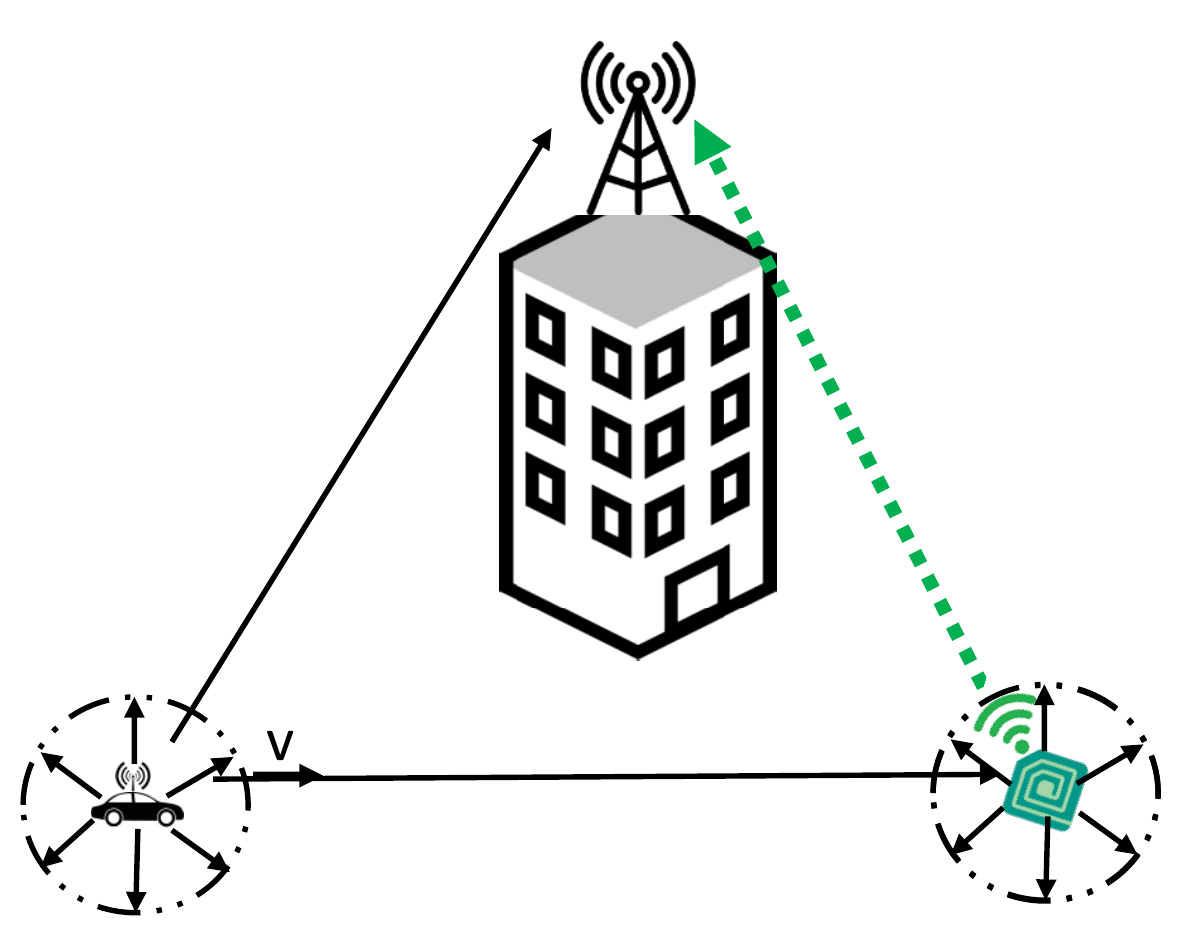}
	\caption{System model for the ambient backscatter setup.}\label{fig:sysmod}
    \end{minipage}
\hfill
    \begin{minipage}[b]{0.48\columnwidth}
        \centering
        \includegraphics [width=\linewidth]{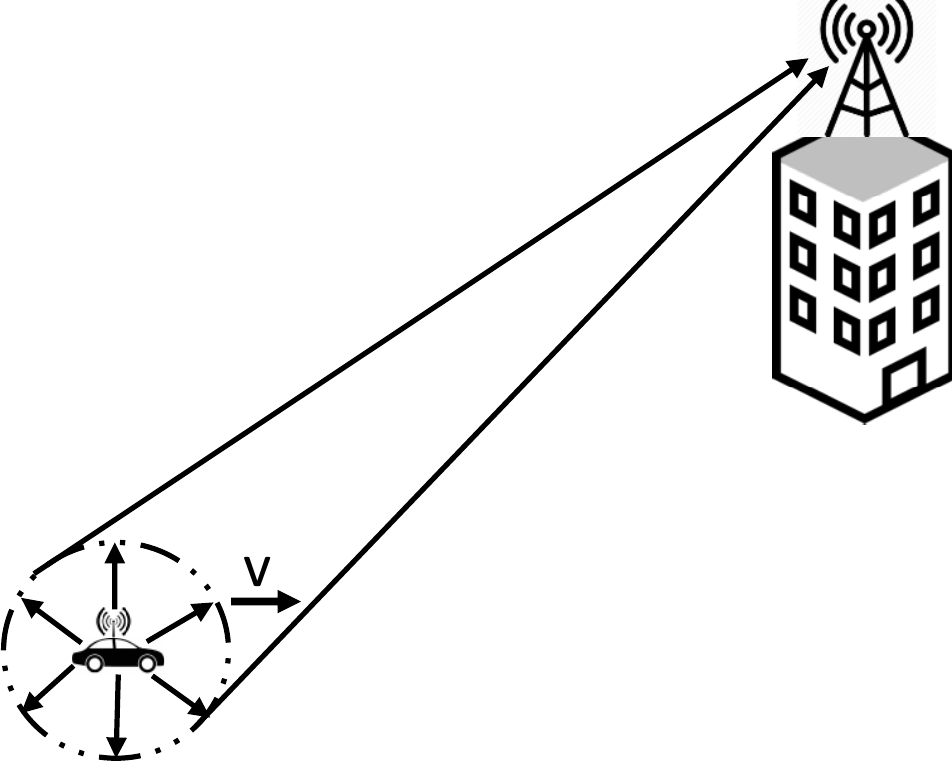}
        \caption{Illustration of the time-selective fading channel.}\label{fig:wchannel}
    \end{minipage}
\end{figure}

The backscatter system in our current setup has three devices: ambient power source (PS), backscatter transmitter (BTx), and receiver (Rx), as illustrated in Fig. \ref{fig:sysmod}. The channel considered in the work is flat Rayleigh faded whose coherence time is of the order of duration of each ambient symbol, with spatial correlation at the Rx. The received signal contains two elements, the DL coming from the ambient PS, and the backscatter link (BL) reflected from the BTx, with their respective AoAs given by $\theta_1$ and $\theta_2$. Both the PS and BTx can be in motion, due to which the channel gain of the three links (including the link from PS to BTx) will be changing with time. As shown in \cite{varshney17}, ambient backscatter can achieve communication with a far away receiver like BS if the PS is not too far and the receiver can find a way to separate the two links, which is the primary motivation for the setup shown in Fig \ref{fig:sysmod}. Emerging applications that motivate the selection of time-selective fading channel for ambient backscatter include smart fabrics where tags/sensors are integrated into garments for monitoring vital signs \cite{wang2017fm}, and sensors deployed on the traffic signs. 
The impulse response of the channel at Rx corresponding to the DL and BL links in terms of the dominant NLOS path and the Rx antenna array response is given as follows \cite{raleigh1994characterization, sayeed2002deconstructing}:
\begin{align}
\mathbf{h}(t) &= \underbrace{\sum\limits_{n=1}^N c_n e^{j \phi_n - j2\pi c\tau_n/\lambda + j 2\pi f_d \cos \psi_n t}}_{ h_0(t)}\mathbf{a}(\theta) \delta\left(t -\bar{\tau}\right),
\end{align}
where the dominant NLOS path can be assumed to be a combination of $N$ independent and non-resolvable sub-paths due to the presence of local scatterers around the transmitter. The $n$th sub-path is characterized by the gain $c_n$, the phase offset $\phi_n$, the time delay $\tau_n$, the maximum Doppler spread (DS) $f_d$, and the angle of departure (AoD) $\psi_n$ at the transmitter, as given in the equation, $\delta$ represents the delta function, and $\bar{\tau}$ is the mean of the individual delays $\tau_n$ of the sub-paths. The remaining parameters $\mathbf{a}(.)$ and $\theta$ are the Rx antenna array response vector, and AoA of the NLOS path, respectively. The phase offset $\phi_n$ of each sub-path is uniformly distributed over $[0, 2\pi)$, and the additional phase offset resulting from the path-delay $\tau_n$ can also be shown to be uniformly distributed over $[0, 2\pi)$ since the frequency of operation is very high \cite[Lemma 4]{dhillon2015wireless}. Applying the central limit theorem (CLT) to the $n$ independent sub-paths, the magnitude of the variable $h_0(t)$ can be shown as Rayleigh distributed. 
This channel environment is illustrated in Fig. \ref{fig:wchannel}. The channel described here is valid when one of the PS or BTx or both are mobile, and the receiver is located above the rooftops (such as BS) resulting in spatial correlation across the antennas. The channel of the PS-BTx link will be similar to $h_0(t)$ with additional DS coming from the local scatterers around BTx. 

The rate at which the coefficient $h_0(t)$ varies is dependent on the maximum Doppler spread $f_d$ and the angular spread $\psi_n$ of the sub-paths at the mobile user. These parameters are large enough in this case due to the movement of the user and the presence of local scatterers, resulting in a fast variation of $h_0(t)$. On the other hand, the array response vector $\mathbf{a}(.)$ depends on the AoA $\theta$ of the NLOS path. The time-scale over which this parameter $\theta$ evolves is several orders of magnitude larger compared to the coherence time of $h_0(t)$, and hence can be tracked by the system. Therefore, while the channel coefficient at the receiver will be changing for each ambient symbol, the angular variation corresponding to AoA of the received signal will not change at the same rate and can be assumed to be constant for few symbol periods. The MA receiver designed in this work will build on this point to improve the BER performance of the system. More information on this property of the fading channels can be found in \cite{adhikary2013joint, adhikary2014joint, adhikary2015massive}.

\begin{remark}\label{rem: chmod2}
 
 The assumption of spatially correlated channel at the Rx is typically valid for a BS located above the rooftops as the angular spread is small in these scenarios. We assume this to be valid for a backscatter device also by considering a single dominant NLOS path. Handling the case of multiple angular paths at the Rx is left as a promising future work. {\cc Further, extension of the non-coherent detection approach proposed in the current work to a frequency-selective channel is another promising area for future investigation.}
\end{remark}


 The auto-correlation function (ACF) of the fading process for the DL and BL links is
\begin{align}
\mathbb{E}_{\theta_n, \tau_n, \bar{\psi}}\left[h_0(t) h_0^*(t+t_d)\right] &= (\sum\limits_{n=1}^N |c_n|^2 ) \mathbb{E}_{\bar{\psi}}[e^{- j 2\pi f_d \cos \bar{\psi} t_d}] =  J_0(2\pi f_d t_d),
\end{align}
where $J_0(.)$ is the zero order Bessel function of the first kind. This result obtained under the assumption of uniformly distributed azimuthal AoD and unit sum energy of the sub-paths is known as Clarke's reference model \cite{stuber2017principles}. Similarly, the ACF for the PS-BTx link is given by $J_0(2\pi f_d t_d) J_0(2\pi a f_d t_d)$, where $a$ is the ratio of the DS at BTx and PS. The Clarke's model cannot be exactly realized in practice, and therefore the Jake's model based on sum of sinusoids is used to generate channel samples that have characteristics similar to the reference model \cite{stuber2017principles}. 
 
\subsubsection*{Autoregressive (AR) modeling of fading channels}

Though Jakes' sum of sinusoids approach to model the temporal-fading process is widely used, it requires large number of sinusoids (and thereby increased complexity) to match the Clarke's reference model and is not mathematically tractable. Hence, this approach is not always convenient to apply for procedures such as channel modeling, estimation and equalization. 
Instead, AR models are used either to decrease the complexity of generating accurate correlated samples of the time-domain fading process or for 
 the derivation of the equalization parameters \cite{baddour2005autoregressive, liu2002space, komninakis2002multi, ghandour2012use}. Therefore, to simplify the analysis, the time-selective fading channel in our work is modeled as an AR process. The correlation matching (CM) criterion of the AR model imposes a condition that the ACF of the approximated process matches the sampled ACF of the Jakes' model. An AR process of order $p$ is given by \cite{baddour2005autoregressive}:
\begin{align}
h[n] &= \sum\limits_{k=1}^{p} a_k h[n-k] + v[n] , \label{eq:arfad1}
\end{align}  
where $v[n]$ is a complex white Gaussian noise process with uncorrelated real and imaginary components. In the case of Rayleigh fading, $v[n]$ has zero mean. The parameters related to the AR model are given by $\{a_1, a_2,..., a_k\}$ and the variance of $v[n]$ by $\sigma_p^2$. The ACF of this approximated process of order $p$ matches exactly with the samples of the desired ACF upto $p$ taps.  The accuracy of this modeling approach using AR process increases with higher order approximations. However, as shown in \cite{wang1996verifying}, the first order AR model obtained by setting $p=1$ is a sufficiently accurate model which can be represented as \cite{liu2002space}:
\begin{align}
	h[n] &=  \rho \, h[n-1] + \sqrt{1-\rho^2}\, g[n], \label{eq:armod1}
\end{align}
where $h[n]$ and $h[n-1]$ are the channel gains in the current and previous time periods, $g[n]$ is the complex white Gaussian noise process with variance $\sigma_h^2$, and $\rho \in [0, 1)$ is the correlation between the fading coefficients of the consecutive symbols. Depending on the link, the correlation factor $\rho$ is given by either $J_0(2\pi f_d T_s)$ or $J_0(2\pi f_d T_s ) J_0(2\pi a f_d T_s)$, where  $T_s$ is the symbol duration. The value of $\rho$ determines the rate at which the current channel coefficient de-correlates across time. The recursive relation in~\eqref{eq:armod1} can be  written in the direct form as:
\begin{align}
	h[n] &= \rho^{n-1} h[1] + \sqrt{1-\rho^2} \left\{\sum\limits_{k=1}^{n-1}\rho^{n-k-1}g[k] \right\}. \label{eq:armod2}
\end{align} 
Note that the modeling of the time-selective fading using the first order AR process in the current work is a good first step, and can be extended to  a higher order AR process in future studies. 

\begin{remark}\label{rem: BFmod}
The time-selective fading implicitly handles the extreme cases of independent fading $(\rho=0)$ and highly correlated fading $(\lim \rho \to 1^{-})$. However, the block fading obtained by configuring $\rho=1$ requires a separate analysis, and will be handled separately in a future work.
\end{remark}

\subsection{Signal Model} \label{sec:SigMod}
In general, the signal scattered from the backscatter device to the receiver is given by \cite{fuschini2008analytical}:
\begin{align}
 r & =  \left( A  - \Gamma \right) s = A s - \Gamma s, \label{eq:totscatEM}
 \end{align}
 where $r$ is the signal at the receiver, $s$ is the signal backcattered at the device, $A$ is the load-independent complex coefficient of the device, and $\Gamma$ is the reflection coefficient of backscatter node at the boundary of the antenna and the circuit. The device modulates the signal by varying the load impedance to change the parameter $\Gamma$ that controls the reflected signal. The first and second terms in~\eqref{eq:totscatEM} correspond to the \emph{structural mode} and \emph{antenna mode} scattering components, respectively. A binary modulation scheme can be implemented by choosing two different values $\Gamma_0$ and $\Gamma_1$. As shown later, non-coherent detection will result in good error performance only for the case of OOK modulation. It is possible to achieve this modulation for antennas with $|A| \le 1$ by designing the appropriate load impedance using only passive components \cite{balanis1997antenna, bletsas2010improving}.
  
Since the data rate of most IoT applications is rather small, it is reasonable to assume that the data rate of backscatter is lower compared to that of the ambient symbols. Under this assumption, a single variable is enough to represent the backscatter data for a signal sample set of size $N$. The signal at the SA receiver is the summation of the direct and backscatter signals, which can be mathematically represented as follows:
\begin{align}
	y[n] &= \underbrace{h_r[n] x[n]}_{\text{direct signal}}+\underbrace{\alpha b \text{ } h_{b}[n] \text{ } h_t[n] x[n]}_{\text{backscatter signal}} + \underbrace{w[n],}_{\text{AWGN}} \label{eq:actmod}
\end{align}
where $x[n]$ is the ambient symbol sequence in complex baseband, $w[n]$ is the additive complex Gaussian noise, $h_r[n], h_b[n]$ and $h_t[n]$ are i.i.d. zero mean complex Gaussian channel coefficients with variance $\sigma_h^2$, $b$ is the backscatter data, $\alpha$ is related to the parameter $\Gamma_1$ of the BTx node. The channel coefficients $h_r[n], h_b[n]$ and $h_t[n]$ are modeled using the AR process of order $1$, each having a different correlation factor given by $\rho_r, \rho_b$, and $\rho_t$, respectively. Since non-coherent detection does not require the CSI, the channel gains $h_r[n], h_b[n]$ and $h_t[n]$ are unknown at the Rx. 
The received signal at the MA receiver with antennas $M_r \ge 2$ is given by:
\begin{align}
	\mathbf{y}[n] &\!\!=\! \!\!\begin{bmatrix}
    y_0[n] \\
    y_1[n]\\
    \vdots\\
    y_{M_r-1}[n]
  \end{bmatrix}\! \!=\! h_r[n] \!\!
    \begin{bmatrix}
    1 \\
    e^{j\phi_1}\\
    \vdots\\
    e^{j(M_r-1)\phi_1}
  \end{bmatrix} \!\!x[n] \!+ \alpha b \, h_{b}[n] h_t[n] \!\!\begin{bmatrix}
    1 \\
    e^{j\phi_2}\\
    \vdots\\
    e^{j(M_r-1)\phi_2}
  \end{bmatrix} \!\!x[n] \!+ \!\!\begin{bmatrix}
    w_0[n] \\
    w_1[n]\\
    \vdots\\
    w_{M_r-1}[n]
  \end{bmatrix}\!,  \label{eq:actmodmo}
\end{align}
where the phase offset $\phi_i$ between consecutive antenna elements for each link is given by $ \frac{2\pi}{\lambda} d \cos \theta_i$ for a linear uniform antenna array. Note that the AoA $\theta_2$ of the BL is independent of the AoA $\theta_1$ of the DL.

The null and alternate hypotheses of the binary hypothesis testing problem are denoted as $\mathcal{H}_0$ and $\mathcal{H}_1$, respectively.  The BTx modulates the backscatter data using the binary on-off keying (OOK) modulation scheme. As is generally the case, the ambient symbol sequence $x[n]$ is assumed to be i.i.d., with unit energy on average. We also assume that the noise energy $\sigma^2_n$, the average channel energy $\sigma^2_h$, and the correlation factors $\rho_r, \rho_b$, and $\rho_t$ are known at the receiver. In fact, they can be perfectly estimated with a long observation interval under the assumption that they remain constant, which is true as they are large-scale parameters. 

\subsubsection*{Test Statistic (TS)}

Due to the reasons outlined in contributions, the receiver architecture is based on the TS of the mean of the received signal samples, unlike the conventional TS of the average energy of the received signal samples. The new TS can be mathematically denoted as:
\begin{align}
Z \mspace{-2mu} &= \frac{1}{N} \sum\limits_{n=1}^{N} y[n] \label{eq:tstat}
\end{align}
It should be noted that derivation of the optimal TS for the time-selective channel is still an open problem. In fact, the optimality of the TS, although important in general, has not really been the main focus of the receiver design for ambient backscatter systems. Some very recent work on the optimal detection and the selection of testing variable for a non-coherent detector under block fading channel can be found in  \cite{darsena2019noncoherent, guruacharya2019optimal}.

\section{Detection at a Single Antenna Receiver} \label{sec:DetectnBERSA}

In this section, we initially derive the growth rate of the expectation and variance of the generalized sum sequence of interest. This result is then used to evaluate the conditional distributions of the signal of the two hypotheses, and ultimately the BER of the SA receiver.

\subsection{Growth Rate of a Generalized Sum Sequence of Interest} \label{sec:concineq}

Consider the general sum sequence $S_N = \sum\limits_{n_1, n_2}  \rho^{|n_1-n_2|} \, x[n_1] x^*[n_2]$, where $m \in \{1,2\}$, defined as the sum of non-i.i.d. RVs, which plays an important role in the signal detection procedure. In particular, the asymptotic property of the sum sequence given by $M_N = \frac{S_N}{N}$ is required to derive the conditional distributions. For this setup, if we can show that the growth rate of both the expectation and variance of $S_N$ is of the order of $N$ (the number of samples), that is sufficient to conclude that the sequence $M_N$ converges to its mean value as the sample size tends to infinity. 
Using the Chebyshev inequality, it is possible to show that this will indeed be the case if the higher order moments of the RV $X$ representing the i.i.d. ambient data sequences $x[n]$ are finite. These conditions on the moments of $x[n]$ might be stronger than necessary but are nevertheless reasonable and assumed here to simplify the derivation. One of the second order moments of the ambient sequences $x[n]$ is the sample energy which is given by:
\begin{align}
\bar{E} &= \mathbb{E} \left[ |X|^2 \right] = \frac {1} {N} \sum\limits_{n=1}^{N} |x[n]|^2 \label{eq: AvgE}.
\end{align}
The result capturing the growth rate of $S_N$ is provided in the following Lemma. Note that one has to be careful in deriving these concentration results since the sum sequence $S_N$ is not composed of i.i.d elements. Please see the proof of the following Lemma for more details. 

\begin{lemma} \label{lem:concineq}
The expectation and variance of the sum sequence $S_N$ both grow asymptotically of the order of $N$, i.e.,  $\mathbb{E}[S_N] = \Theta(N)$ and ${\rm Var}[S_N] = \Theta(N)$, where $f(x) \mspace{-3mu}=\mspace{-3mu} \Theta (g(x))$ means that $f(x)$ is asymptotically bounded both from above and below by $g(x)$. As a consequence, the sequence $M_N$ concentrates around $\mathbb{E} [M_N]$ when $N\rightarrow \infty$, where
\begin{align}
\mathbb{E}\left[ M_N \right] &=  \mathbb{E} \left[ |X|^2 \right] + \frac{2\rho}{1-\rho} \left(1 - \frac{1-\rho^N}{N(1-\rho)} \right) |\mathbb{E}[X]|^2. \label{eq:ExpMn}
\end{align} 
\end{lemma}

\begin{IEEEproof}
	See Appendix~\ref{app:concineq}.
\end{IEEEproof}

\begin{figure}
    \centering
    \begin{subfigure}[b]{0.45\textwidth}
        \centering
        \includegraphics [width=\linewidth]{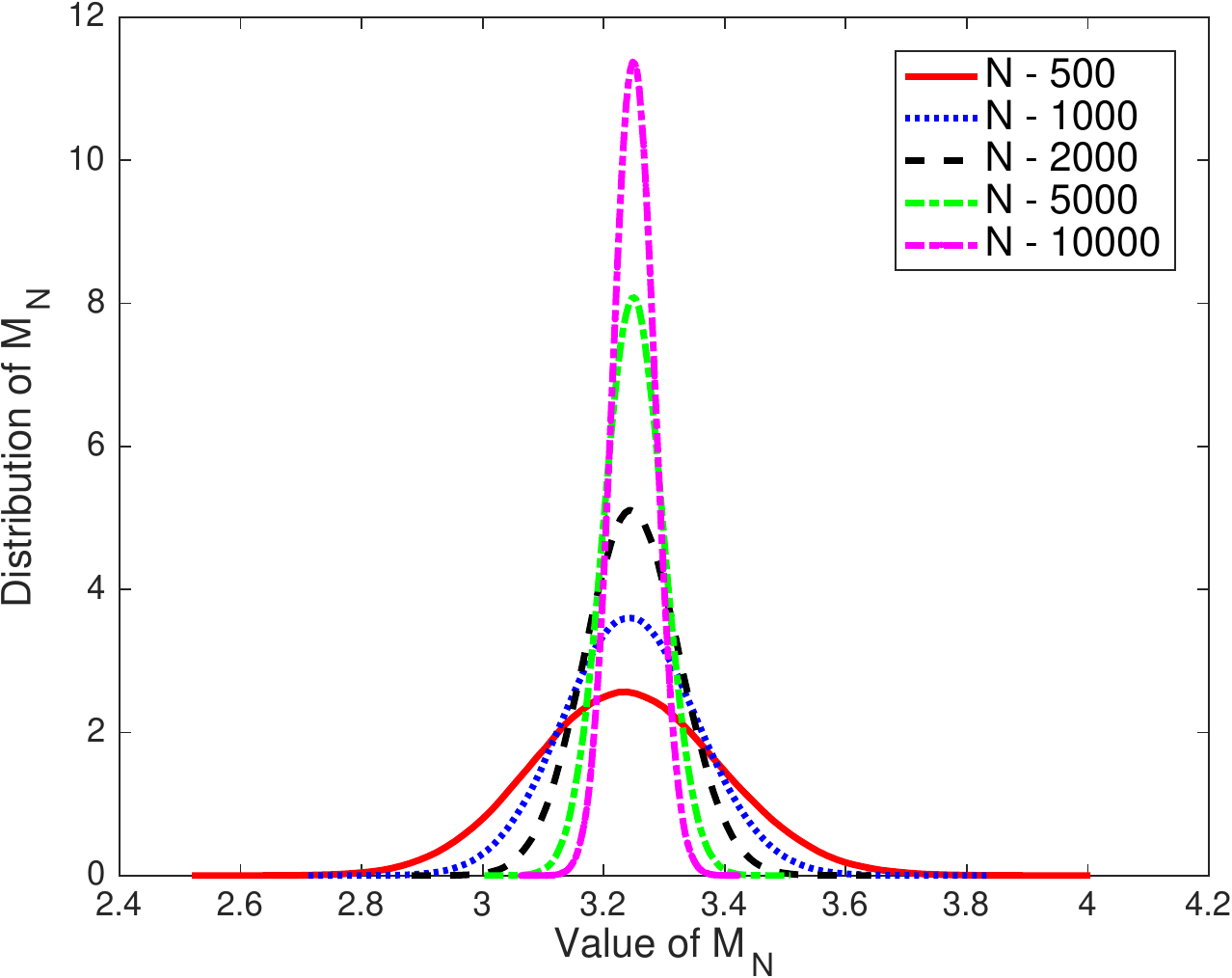}
         \caption{}\label{fig:var_conc_Z_H0}
    \end{subfigure}
    ~ 
    \begin{subfigure}[b]{0.45\textwidth}
        \centering
        \includegraphics [width=\linewidth]{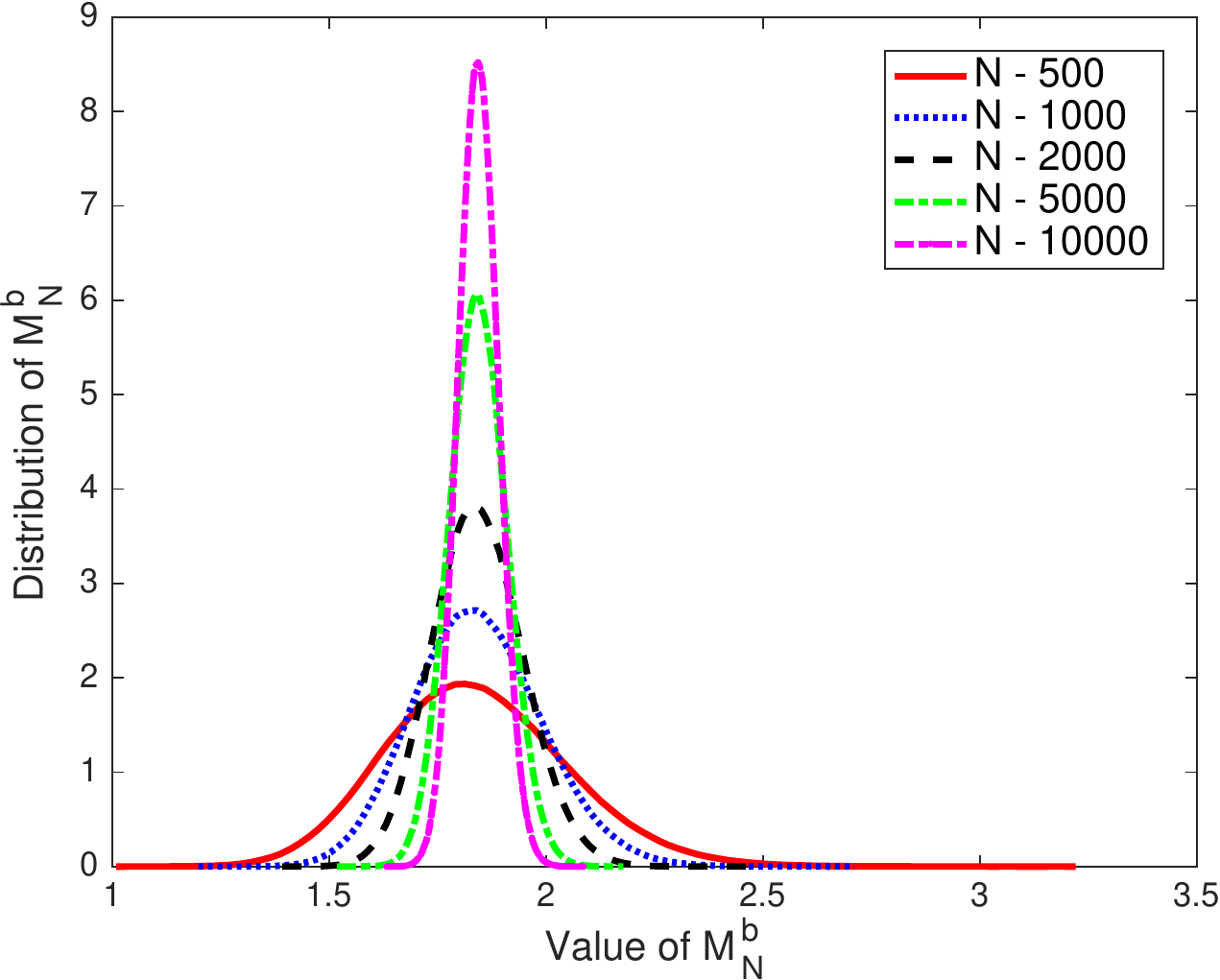}
	\caption{}\label{fig:var_conc_Zb_H1}
    \end{subfigure}
    \caption{ Probability density functions of (a) $M_N$ and (b) $M^b_N$ for varying $N$ with $\rho=0.6$.} 
\end{figure}

The analysis related to Lemma \ref{lem:concineq} on the asymptotic growth rate of $S_N$ is discussed now by plotting the simulation results. The plots for the distributions of $M_N$ and $M^b_N$ with increasing sample size $N$ are shown in Figs. \ref{fig:var_conc_Z_H0} and \ref{fig:var_conc_Zb_H1}, where it can be observed that the mean values remain constant while their variances decrease as the signal sample size increases.

\subsection{Conditional Distributions of the Signal} \label{sec:ConDistM1}

The null and alternate hypotheses $\mathcal{H}_0$ and  $\mathcal{H}_1$ correspond to the scenarios of the transmitted backscatter data $b \equiv 0$ and $b \equiv 1$, respectively. 

\subsubsection{Null Hypothesis $\mathcal{H}_0$}

For the AR process used in this paper to model the time-selective fading, the channel gain evolves with time according to~\eqref{eq:armod1}, which can be described as a weighted average of the previous channel gain and a new variable. Due to this dependence of the current channel gain on the previous gains, the received signal samples are correlated, and hence the co-variances of the samples are non-zero in general. As a consequence, both the variances and co-variances of the signal has to be evaluated to derive the variance of the the mean received signal $Z$. 
 We transform the expression for each received signal sample into a sum representation of independent RVs. This will simplify the evaluation of both the variance of each signal sample and the subsequent evaluation of the variance of $Z$. This can be represented as following: 
\begin{align}
y[n] &= h_r[n] x[n] + w[n] = \left( \rho_r^{n-1} h_r[1] + \sqrt{1-\rho_r^2} \left\{\sum\limits_{k=1}^{n-1}\rho_r^{n-k-1}g_r[k] \right\} \right) x[n] + w[n], \label{eq:nullhypeq}
\end{align}
where the channel gain $h_r[1]$ in the first time slot of a window 
can be independently configured.

\begin{lemma} \label{lem:CondPDFH0M1}
The probability density function (PDF) of $Z$ conditioned on $\mathcal{H}_0$ is given by 
\begin{align}	
	\mathcal{H}_0 &: Z \sim  \mathcal{CN} \left(0, {\rm Var}^{\rm SA}_0 \right)  \label{eq:condpdf1M1lem},
\end{align}
where ${\rm Var}^{\rm SA}_0 \mspace{-3mu}=\mspace{-3mu} \frac{\sigma_h^2 \mathbb{E} \left[ |X|^2 \right] + \sigma_h^2 \frac{2\rho_r}{1-\rho_r} \left( 1 - \frac{1-\rho_r^N}{N(1-\rho_r)} \right) |\mathbb{E} \left[ X \right]|^2 + \sigma_n^2}{N}$ is the conditional variance of $\mathcal{H}_0$.
\end{lemma}

\begin{IEEEproof}
	See Appendix~\ref{app:CondPDFH0M1}.
\end{IEEEproof}

%

\subsubsection{Alternate Hypothesis $\mathcal{H}_1$}

The received signal for a sample $n$, where $1 \le n \le N$, under the alternate hypothesis $\mathcal{H}_1$ is given by:
\begin{align}
y[n] &=  h_r[n] x[n] + \underbrace{\alpha h_b[n] h_t[n] x[n]}_{y_b[n]} + w[n], \label{eq:althypeq}
\end{align}
where $ h_r[n], h_b[n]$ and $h_t[n]$ are the fading gains following the process defined by~\eqref{eq:armod2}. Unlike the case of $\mathcal{H}_0$, further work is needed to derive the distribution for $\mathcal{H}_1$ since the conditional distribution of each sample is not complex Gaussian anymore. However, we preserve the Gaussian property of the samples by further conditioning on $h_b[n]$ and show that this conditional distribution asymptotically matches the true distribution. Only the distribution corresponding to $y_b[n]$ is needed to be derived and the sequence $M^b_N = \sum\limits_{n_1, n_2}  \rho_t^{|n_1-n_2|} \, h_b[n_1] h_b^*[n_2] x[n_1] x_2^*[n_2]$ related to $y_b[n]$ is the corresponding parameter of $\mathcal{H}_1$, similar to $M_N$ of $\mathcal{H}_0$. The following Lemma captures this analysis on the conditional distribution of $\mathcal{H}_1$.

\begin{lemma} \label{lem:CondPDFH1M1}
The PDF of $Z$ conditioned on $\mathcal{H}_1$ is given by 
\begin{align}	
	\mathcal{H}_1 &: Z \sim  \mathcal{CN} \left(0,{\rm Var}^{\rm SA}_1\right)  \label{eq:condpdf2M1lem},
\end{align}
where ${\rm Var}^{\rm SA}_1 = \frac{\sigma_h^2 (1 + |\alpha|^2  \sigma_h^2) \mathbb{E} \left[ |X|^2 \right] + \sigma_h^2 \left[\frac{2\rho_r}{1-\rho_r} \left( 1 - \frac{1-\rho_r^N}{N(1-\rho_r)} \right)
	 +  |\alpha|^2  \sigma_h^2 \frac{2\rho_t \rho_b}{1-\rho_t \rho_b} \left( 1 - \frac{1-\rho_t^N \rho_b^{N}}{N(1-\rho_t \rho_b)} \right) \right] |\mathbb{E} \left[ X \right]|^2  + \sigma_n^2}{N} $.
\end{lemma}

\begin{IEEEproof}
	See Appendix~\ref{app:CondPDFH1M1}.
\end{IEEEproof}
The results are valid for all $\rho_r, \rho_b$ and $\rho_t \in [0, 1)$,  and the special case of independent fading analyzed in the conference version \cite{devineni19noncoherent} can be obtained by configuring $\rho_r, \rho_b$ and $\rho_t$ all to zero.

\subsection{Bit Error Rate} \label{sec:DetBERSO}

From the conditional distribution analysis, we see that the PDFs of the two hypotheses have same mean but different variances, which are compared to obtain the optimal detection threshold.

\begin{theorem} \label{thm:BERExpSOM1}
The average BER of a SA receiver is given by
\begin{gather}
P_{\rm SA}(e) = \frac{1}{2} - \frac{1}{2}  e^{-\frac{T_{\rm SA}}{{\rm Var}^{\rm SA}_1}} + \frac{1}{2}  e^{-\frac{T_{\rm SA}}{{\rm Var}^{\rm SA}_0}}, \label{eq:avgberSA}
\end{gather}
where $T_{\rm SA} = \ln\left( \frac{{\rm Var}^{\rm SA}_1}{{\rm Var}^{\rm SA}_0}\right)\frac{{\rm Var}^{\rm SA}_1{\rm Var}^{\rm SA}_0}{{\rm Var}^{\rm SA}_1-{\rm Var}^{\rm SA}_0} $ is the optimal detection threshold.
\end{theorem}

\begin{IEEEproof}
	See Appendix~\ref{app:BERExpSOM1}.
\end{IEEEproof} 

\subsubsection*{\textbf{Asymptotic analysis}}
The ratio of the variances of $\mathcal{H}_0$ and $\mathcal{H}_1$ of the SA receiver is:
\begin{align}
K =  \frac{{\rm Var}^{\rm SA}_1}{{\rm Var}^{\rm SA}_0} 
&= 1 + \frac{ |\alpha|^2 \sigma_h^4 \left\{ 1 +  \frac{2\rho_t \rho_b}{1-\rho_t \rho_b} \left( 1 - \frac{1-\rho_t^N \rho_b^{N}}{N(1-\rho_t \rho_b)} \right) \frac{ |\mathbb{E} \left[ X \right]|^2}{\mathbb{E} \left[ |X|^2 \right]}  \right\}}{\sigma_h^2 \left\{1 +  \frac{2\rho_r}{1-\rho_r} \left( 1 - \frac{1-\rho_r^N}{N(1-\rho_r)} \right) \frac{ |\mathbb{E} \left[ X \right]|^2}{\mathbb{E} \left[ |X|^2 \right]}  \right\} + {\rm SNR}^{-1}}. 
\end{align}
The asymptotic average BER can be simplified as follows:
\begin{align}
P_{\rm SA}^{\rm asym}(e) 
&\stackrel{(a)}{=} \lim_{{\rm SNR} \to \infty} \frac{1}{2}  (1 - K^{\frac{-1}{K-1}} + \frac{1}{K}^{\frac{1}{1-\frac{1}{K}}} ) \stackrel{(b)}{=} \frac{1}{2}  (1 - K_{\infty}^{\frac{-1}{K_{\infty}-1}} + \frac{1}{K_{\infty}}^{\frac{1}{1-\frac{1}{K_{\infty}}}} ), \label{eq:asymBERM0SA}
\end{align}
where (a) results from the substitution of the expression for $T_{\rm SA}$ and replacing the ratio $\frac{{\rm Var}^{\rm SA}_1}{{\rm Var}^{\rm SA}_0}$ with $K$ defined earlier, and (b) follows from the substitution of $K$ with $K_{\infty}$ obtained as ${\rm SNR} \to \infty$.

\begin{remark}\label{rem: condpdf}

 Clearly, the BER expressions under the new receiver architecture are independent of $N$ when the expectation $\mathbb{E} \left[ X \right]$ of the ambient data sequence is zero and/or the time-domain fading is uncorrelated (all the $\rho$'s equal $0$). Furthermore, the asymptotic BER value, with respect to the increasing SNR, reaches an error floor instead of converging to zero. This error floor is numerically demonstrated later in Fig. \ref{fig:BERSOMO}. This necessitates the need to develop better techniques to decode data in a time-selective channel, which takes us to the next main contribution.
%
\end{remark}

\section{Detection at a Multi-Antenna Receiver} \label{sec:DetectnBERMA}

\subsection{Effective Signal and Antenna Gain} \label{sec:CondPDFMA}
The main reason for the poor BER performance of the SA receiver is the presence of the DL from the ambient PS, which only acts as an interference since it does not carry any backscatter data. 
The signals impinging on the neighboring antenna elements are phase shifted versions of the signal at the first antenna in addition to the independent additive noise. Observe that the phase offset of the BL is independent of the phase offset of the DL. The interference of the DL can be canceled by reversing the DL phase offset at each antenna starting from the second element, and subtracting the resultant signal with that at the first antenna, as given below:
\begin{align}
	\tilde{\mathbf{y}}[n] = \begin{bmatrix}
    e^{- j\phi_1} y_1[n] - y_0[n]\\
    \vdots\\
     e^{- j(M_r-1)\phi_1} y_{M_r-1}[n]- y_0[n]
  \end{bmatrix} &= \tilde{\mathbf{a}} \, \alpha b \, h_{b}[n] h_t[n]  x[n] + \tilde{\mathbf{w}}[n],  \label{eq:effsignal}
\end{align}
where the effective antenna array and noise vectors $\tilde{\mathbf{a}}$ and $\tilde{\mathbf{w}}[n]$, respectively, are given by:
\begin{align}
\tilde{\mathbf{a}} 
&= \begin{bmatrix}
     2 \sin (\frac{\phi_2-\phi_1}{2}) e^{j(\frac{\phi_2-\phi_1}{2})}  \\
    \vdots\\
    2 \sin (M_r-1)(\frac{\phi_2-\phi_1}{2}) e^{j(M_r-1)(\frac{\phi_2-\phi_1}{2})} 
  \end{bmatrix}, 
   \tilde{\mathbf{w}}[n] &= \begin{bmatrix}
    e^{- j\phi_1} w_1[n] - w_0[n] \\
    \vdots\\
    e^{- j(M_r-1)\phi_1} w_{M_r-1}[n] - w_0[n] 
  \end{bmatrix}.
\end{align}
The covariance matrix of the resultant noise vector $\tilde{\mathbf{w}}[n]$ is given by:
\begin{align}
\mathbf{K_{\tilde{W}}} = \sigma_n^2 \mathbf{\hat{K}_{\tilde{W}}}, \quad \text{where } \mathbf{\hat{K}_{\tilde{W}}}  = \begin{bmatrix}
   2 & 1 & \ldots & 1  \\
   \vdots & \vdots & \ddots & \vdots\\
   1 & 1 & \ldots & 2
  \end{bmatrix},
\end{align}
which means that the resultant noise after the DL cancellation is correlated. The vector detection problem can be converted to scalar detection by appropriately designing the weight vector. The effective scalar signal samples for the averaging operation can be obtained by the following steps: 1) Whiten the additive noise with the linear transformation $\mathbf{\hat{K}_{\tilde{W}}^{-\frac{1}{2}}}$, and 2) Project the output of the first step along the direction of the resultant antenna array response $\mathbf{\hat{K}_{\tilde{W}}^{-\frac{1}{2}}} \tilde{\mathbf{a}} $.  The combined weight vector of the two operations is $\mathbf{r} =  \frac{ \mathbf{\hat{K}_{\tilde{W}}^{-1}} \tilde{\mathbf{a}}}{|\mathbf{\hat{K}_{\tilde{W}}^{-\frac{1}{2}}} \tilde{\mathbf{a}}|}$, and the effective signal after these steps is:
\begin{align}
 y_{\rm eff}[n] &= \mathbf{r}^* \tilde{\mathbf{y}}[n]  = \frac{ \tilde{\mathbf{a}}^* \mathbf{\hat{K}_{\tilde{W}}^{-1} \tilde{\mathbf{a}}}}{|\mathbf{\hat{K}_{\tilde{W}}^{-\frac{1}{2}}} \tilde{\mathbf{a}}|} \alpha b \, h_{b}[n] h_t[n]  x[n] + \frac{ \tilde{\mathbf{a}}^* \mathbf{\hat{K}_{\tilde{W}}^{-1}} }{|\mathbf{\hat{K}_{\tilde{W}}^{-\frac{1}{2}}} \tilde{\mathbf{a}}|} \tilde{\mathbf{w}}[n] \label{eq:effsignalMr}.
\end{align}
Hence, the gain in the average signal power with multiple antennas is $\tilde{\mathbf{a}}^* \mathbf{\hat{K}_{\tilde{W}}^{-1}} \tilde{\mathbf{a}}$, while the noise power remains at $\sigma_n^2$. Therefore, the antenna gain (SNR) due to multiple antennas is given by $\tilde{\mathbf{a}}^* \mathbf{\hat{K}_{\tilde{W}}^{-1}} \tilde{\mathbf{a}}$. This procedure to generate the scalar sample $y_{\rm eff}[n]$ maximizes the SNR of the signal. In addition the resultant sample $y_{\rm eff}[n]$ is a sufficient statistic for the detection procedure that follows. It can be further shown that this procedure also minimizes the mean square error for the signal estimation, and is hence known as the linear minimum mean
squared error estimation (MMSE) \cite{dtse}. 
 The phase-offset components $e^{j\phi_1}$ and $e^{j\phi_2}$ of the two links can be estimated from the received signal by formulating a parameter estimation problem. However, this is beyond the scope of the current work, and hence they are assumed to be perfectly known at the receiver. The sample average given by $Z = \frac{1}{N} \sum\limits_{n=1}^{N} y_{\rm eff}[n]$ is used as the new test statistic for detection. 

\begin{lemma} \label{lem:SNRgain}
The antenna (SNR) gain $ G =\tilde{\mathbf{a}}^* \mathbf{\hat{K}_{\tilde{W}}^{-1}} \tilde{\mathbf{a}}$ of the MA receiver is given by:
\begin{align}
G &= M_r\!-\!\frac{1}{M_r} \!- \!\frac{2}{M_r} \frac{\sin  \left((M_r\!-\!1)\frac{\phi_2-\phi_1}{2} \right) }{\sin \left(\frac{\phi_2-\phi_1}{2}  \right)}  \cos \!\left(\! \frac{M_r}{2} (\phi_2-\phi_1)\!\right) \!  - \!\frac{1}{M_r} \frac{\sin^2  \left((M_r\!-\!1)\frac{\phi_2-\phi_1}{2} \right) }{\sin^2 \left(\frac{\phi_2-\phi_1}{2} \right)}.
\end{align}
\end{lemma}
\begin{IEEEproof}
	See Appendix~\ref{app:SNRgain}.
\end{IEEEproof}
For notational simplicity, the antenna gain is represented as a single variable $G$ without any input arguments even though it is a function of the two phase offsets (and hence the AoAs).

\begin{remark}\label{rem:genmrspecial}
 The antenna gain of a dual-antenna Rx ($M_r=2$) simplifies to $G = 2 \sin^2 \left(\frac{\phi_2-\phi_1}{2} \right)$, which is zero when the AoAs of the DL and BL links are almost the same. On the other hand, the antenna gain $G$ for a Rx with $M_r>2$ equals $(1-\frac{1}{M_r})(M_r-2)$, which is non-zero even when the two AoAs are almost the same. Hence, additional angular resolution is obtained with $M_r>2$, which is useful for the applications where the AoAs of the DL and BL links are similar. 
\end{remark}

\subsection{Conditional Distributions of the Effective Signal and Bit Error Rate} \label{sec:condDistDetBERMO}

Now, we derive the conditional distributions of the effective signal derived in~\eqref{eq:effsignalMr}, and then use them to evaluate the average BER of the MA receiver.

\begin{lemma} \label{lem:CondPDFMA}
The conditional PDFs of $Z$ for the two hypotheses $\mathcal{H}_0$ and $\mathcal{H}_1$ are given by
\begin{align}	
	\mathcal{H}_i: 
Z \sim \mathcal{CN} \left(0,{\rm Var}^{\rm MA}_i\right), \label{eq:MOHi} 
\end{align}
where ${\rm Var}^{\rm MA}_0 = \frac{ \sigma_n^2}{N}$ and ${\rm Var}^{\rm MA}_1 = \frac{ G |\alpha|^2 \sigma_h^4 \left\{ \mathbb{E} \left[ |X|^2 \right] +  \frac{2\rho_t \rho_b}{1-\rho_t \rho_b} \left( 1 - \frac{1-\rho_t^{N}\rho_b^{N}}{N(1-\rho_t \rho_b)} \right) |\mathbb{E} \left[ X \right]|^2  \right\}+ \sigma_n^2}{N}$. 
\end{lemma}

\begin{IEEEproof}
	See Appendix~\ref{app:CondPDFMA}.
\end{IEEEproof}

\begin{theorem} \label{thm:BERExpMO}
The average BER of the MA receiver is  given by:
\begin{align}
&P_{\rm MA}(e) = \int_{-\pi}^{\pi} \int_{-\pi}^{\pi} \frac{1}{2\pi} \times \frac{1}{2\pi} \times \frac{1}{2}  \left(1 - e^{{-\frac{T_{\rm MA}}{{\rm Var}^{\rm MA}_1}}} + e^{{-\frac{T_{\rm MA}}{{\rm Var}^{\rm MA}_0}}}\right) \mathrm{d} \theta_1 \mathrm{d} \theta_2,
\end{align}
where $T_{\rm MA} = \ln\left( \frac{{\rm Var}^{\rm MA}_1}{{\rm Var}^{\rm MA}_0}\right)\frac{{\rm Var}^{\rm MA}_1{\rm Var}^{\rm MA}_0}{{\rm Var}^{\rm MA}_1-{\rm Var}^{\rm MA}_0}$ is the optimal detection threshold.
\end{theorem}

\begin{IEEEproof}
	See Appendix~\ref{app:BERExpMO}.
\end{IEEEproof} 
{\cc It should be noted that the SA scenario is not exactly a special case of the MA scenario, even though there are similarities in the non-coherent detection approach and the subsequent bit error rate evaluation of the two receivers. Mainly, the additional operation of the DL interference cancellation in the MA scenario results in an effective antenna array vector and correlated additive noise, which necessitates the handling of the MA receiver separately from the SA receiver.}

\subsubsection*{\textbf{Asymptotic analysis}}
The ratio of the variances of $\mathcal{H}_0$ and $\mathcal{H}_1$ of the MA receiver is:
\begin{align}
K &=  \frac{{\rm Var}^{\rm MA}_1}{{\rm Var}^{\rm MA}_0} 
= 1 + G |\alpha|^2 \sigma_h^4  \left\{ 1 +   \frac{2\rho_t \rho_b}{1-\rho_t \rho_b} \left( 1 - \frac{1-\rho_t^{N}\rho_b^{N}}{N(1-\rho_t \rho_b)} \right) \frac{ |\mathbb{E} \left[ X \right]|^2}{\mathbb{E} \left[ |X|^2 \right]} \right\} {\rm SNR} .\label{eq:KeqMA}
\end{align}
From this, the asymptotic conditional BER of the MA receiver as ${\rm SNR} \to \infty$ can be derived as:
\begin{align}
P_{\rm MA}^{\rm asym}(e|\phi_1, \phi_2) &= \frac{1}{2}  (1 - e^{{-\frac{T_{\rm MA}}{{\rm Var}^{\rm MA}_1}}} + e^{{-\frac{T_{\rm MA}}{{\rm Var}^{\rm MA}_0}}}) \stackrel{(a)}{=} \frac{1}{2}  (1 - K^{\frac{-1}{K-1}} + \frac{1}{K}^{\frac{1}{1-\frac{1}{K}}}) \stackrel{(b)}{=} 0,  \label{eq:asymBERM0MA}
\end{align}
where (a) results from substituting the expression for $T_{\rm MA}$, and replacing $\frac{{\rm Var}^{\rm MA}_1}{{\rm Var}^{\rm MA}_0}$ with $K$ defined in~\eqref{eq:KeqMA}, and (b) follows from the standard limit $\lim\limits_{x \to \infty} (x)^{-1/x-1} = 1$, and $\frac{1}{K} \to 0$ as ${\rm SNR} \to \infty$.
It should be noted that the asymptotic value of $K$ when $N \to \infty$ is non-zero. Hence, the BER does not converge to $0.5$ as $N \to \infty$ even though the individual variances converge to zero.
\begin{remark}\label{rem: averageber}
In case of fast-fading, where the fading gains are independent across the ambient symbols, the average BER is only dependent on the expected value of the energy of the ambient symbol. This special case concurs with our analysis in \cite{devineni19noncoherent}. Alternatively, if the mean value of the ambient symbol is zero (which is the case for most of the modulation schemes), then again the average BER is only dependent on the expected value of the energy. Lastly, it can be inferred from the BER expression that the average BER is an increasing function of the correlation factor.
\end{remark}

\section{Receiver Synchronization and Parameter Estimation} \label{sec:SynchAmB}

\subsection{Delay Parameters}

In this section, we discuss receiver synchronization in ambient backscatter, which is an important ingredient of the proposed system design. First, we briefly mention the parameters to be estimated, and then either provide an analysis of the impact of incorrect estimation of the parameter on the detection performance or provide a procedure to estimate the parameter. In general, the estimation of both the timing delay and the carrier phase offset is necessary in a communication system. In our setup, however, carrier phase estimation is not required since non-coherent detection is employed. As the symbol duration of the backscatter data is larger than that of the ambient data, it is not required to perform symbol synchronization at the backscatter device. Therefore, the symbol timing recovery at the receiver is our main concern. The parameters $T_a$ and $N$ represent the duration of the ambient symbol and the sample window size at the receiver, respectively, which are assumed to be known {\em a priori}. The duration of the backscatter symbol $T_b$ is related to the above two parameters as $T_b= N T_a$. Due to the architecture adopted at the receiver, it needs to estimate the following parameters: (i) the timing delay $\tau \in [0, T_a)$ of the ambient data to obtain signal samples correctly, and (ii) the sample number $k \in \{0, 1, 2, \cdots, N-1\}$ to reset the counter of the signal sample window. The estimation of the delay $\tau$ for time-selective fading channels is a well-studied topic, where correlation-based techniques are widely used to solve the ML estimation problem \cite{mengali2013synchronization}. However, before going into those details, it would be worthwhile to investigate how significant would be the impact of incorrect estimation of the delay (given by $\hat{\tau}$) on the achievable BER. 
For the purpose of exposition, we assume that the pulse shape of the ambient symbols is rectangular, and hence the matched filter pulse is also rectangular. Due to the mismatch of the estimated delay, the discrete samples obtained at the SA receiver after the matched filtering can be represented as: 
\begin{align*}
y[n] &=  \frac{\Delta \tau}{T_a} e^{j\phi_r} h_r[n-1] x[n-1] +  \frac{T_a-\Delta \tau}{T_a} e^{j\phi_r} h_r[n] x[n] \nonumber\\
& \quad \quad \quad + \frac{\Delta \tau}{T_a} e^{j\phi_b} \alpha h_t[n-1] h_b[n-1] x[n-1] + e^{j\phi_r} \alpha h_t[n] h_b[n] \frac{T_a-\Delta \tau}{T_a} x[n] + w[n],
\end{align*}
where $\Delta \tau = \tau - \hat{\tau} \in [0, T_a)$ equals the difference of the actual and the estimated path delays. 

For a window of samples, the average of the samples will simplify as follows:
\begin{align}
Z &= \frac{1}{N} \sum\limits_{n=1}^{N} y[n] =  \frac{e^{j\phi_r}}{N}  ( \frac{\Delta \tau}{T_a} h_r[-1] x[-1] + \sum\limits_{n=1}^{N-1} h_r[n] x[n] +  \frac{T_a-\Delta \tau}{T_a} h_r[N] x[N] ) \nonumber\\
& \quad + \frac{e^{j\phi_b} \alpha } {N} ( \frac{\Delta \tau}{T_a} h_t[-1] h_r[-1] x[-1] + \sum\limits_{n=1}^{N-1} h_t[n] h_b[n] x[n] +  \frac{T_a-\Delta \tau}{T_a} h_t[N] h_b[N] x[N] ). \label{eq:timingavg}
\end{align}
From~\eqref{eq:timingavg}, it is clear that the impact of the timing recovery error on $Z$ (and hence on the BER) will be negligible. In fact, due to the linear averaging operation of the new architecture, the receiver is robust to synchronization errors, and it does not require the estimation of delay $\tau$. 


For the other parameter  of interest $k$, a procedure for estimation is provided. Suppose that backscatter device sends a preamble sequence of alternating bits $1010 \cdots10$ (of length $N_b$), and the index $k$ represents the delay reset of the counter corresponding to the window of signal samples. It should be noted here that the  alternating bit sequences are commonly used in conventional networks for clock and frame synchronization, e.g., see \cite{pahlavan2005wireless}. The sample mean corresponding to backscatter symbols of $'0'$ and $'1'$ taken with a delay $l$ are denoted as $Z^l_0$ and $Z^l_1$, respectively. 
For the purpose of exposition, assume that the delay $k$ is zero. When the sampling window is aligned properly, the energy of the average of the samples $|Z_0^0|^2$ and $|Z_1^0|^2$  corresponding to symbols $'0'$ and $'1'$ can be approximated by ${\rm Var}_0^{\rm SA}$ and ${\rm Var}_1^{\rm SA}$, respectively. Since, the received signal corresponding to symbol  $'1'$ has both the DL and BL links, ${\rm Var}_1^{\rm SA}$ is higher compared to ${\rm Var}_0^{\rm SA}$ resulting in the ratio $\frac{|Z^0_1|}{|Z^0_0|} = C >1$. When the sampling window is misaligned by exactly half the window size $N$, then both $Z^{N/2}_0$ and $Z^{N/2}_1$ contain equal number of ambient symbols that correspond to backscatter symbols $'1'$ and $'0'$, resulting in the ratio $\frac{|Z^{N/2}_1|}{|Z^{N/2}_0|} =1$. In fact, the ratio $\frac{|Z^l_1|}{|Z^l_0|}$ for a general delay $l$ will lie in the interval $(1, C)$. From this, one can conclude that $\frac{|Z^l_1|}{|Z^l_0|}$ is maximized when the sample window is aligned to the delay $k$, and therefore the problem of estimating the parameter $k$ can be formulated as following:
\begin{align}
\hat{k} = \underset{l \in \{0, 1, \cdots N-1\}}{\mathrm{arg \,max}} \frac{|Z^l_1|}{|Z^l_0|}.
\end{align}

{ \cc
\subsection{Correlation Factor and Phase Offset Inversion Parameters} \label{sec:PhOffInv}

Supposing that the delay $k$ is perfectly estimated in the synchronization module, consider the consecutive samples $y_0^k[n]$ and $y_0^k[n+1]$ corresponding to the preamble bit $0$ at the SA receiver. Taking cross-correlation of the two signals, the DL correlation factor $\rho_r$ can be evaluated as:
\begin{align}
&E[y_0^k[n] (y_0^k[n+1])^*] =  \mathbb{E} \left[(h_r[n] x[n]+w[n]) ((h_r^*[n+1] x^*[n+1]+w^*[n+1]) \right] \nonumber\\
&= \rho_r \mathbb{E}\left[ |h_r[n]|^2 \right] |\mathbb{E} \left[ X \right]\!|^2 = \rho_r \sigma_h^2 |\mathbb{E} \left[ X \right]\!|^2 
\implies  \rho_r = \frac{E[y_0^k[n] (y_0^k[n+1])^*]}{\sigma_h^2 |\mathbb{E} \left[ X \right]\!|^2}.
\end{align}
Similarly, the combined correlation factor $\rho_t \rho_b$ of the BL can be evaluated as follows:
\begin{align}  
\rho_t \rho_b &= \frac{E[y_1^k[n] (y_1^k[n+1])^*] - E[y_0^k[n] (y_0^k[n+1])^*]}{|\alpha|^2 \sigma_h^4 |\mathbb{E} \left[ X \right]\!|^2}.
\end{align}

Now, consider the parameters $\mathbf{Z}^k_0$ and $\mathbf{Z}^k_1$ of the MA receiver for deriving the phase offset inversion components $e^{-j\phi_1}$ and $e^{-j\phi_2}$. Next, we provide a method to determine $e^{-j\phi_1}$ of the DL at the receiver. The samples corresponding to the preamble bit $0$ at the consecutive antenna elements $m$ and $m+1$ of the MA receiver are given by:
\begin{align*}
	\begin{bmatrix}
    y^k_{0,m}[n] \\
    y^k_{0,m+1}[n]
  \end{bmatrix} &= h_r[n] e^{j m\phi_1}
    \begin{bmatrix}
    1 \\
    e^{j\phi_1}
  \end{bmatrix} x[n] + \begin{bmatrix}
    w_m[n] \\
    w_{m+1}[n]
  \end{bmatrix}.
\end{align*}
 Taking the mean over samples for each preamble symbol of value $0$ will result in:
\begin{align*}
	\mathbf{Z}^k_0 \!=\!  \sum\limits_{n=1}^{N} \frac{ \mathbf{y}^k_0[n]}{N} \!&=\! \! \begin{bmatrix}
    \sum\limits_{n=1}^{N} \frac{ y^k_{0,m}[n] }{N} \\
    \sum\limits_{n=1}^{N} \frac{ y^k_{0,m
    +1}[n] }{N}
  \end{bmatrix} \!\!=\! \sum\limits_{n=1}^{N} \frac{ h_r[n] e^{j m\phi_1}  x[n] }{N} 
   \! \begin{bmatrix}
    1 \\
    e^{j\phi_1}
  \end{bmatrix} \!\! +\! \!\begin{bmatrix}
    \sum\limits_{n=1}^{N} \frac{ w_m[n] }{N } \\
    \sum\limits_{n=1}^{N} \frac{ w_{m+1}[n] }{N}
  \end{bmatrix} \!\!  =\! c_0 \!\begin{bmatrix}
    1 \\
    e^{j\phi_1}
  \end{bmatrix} \!\!+\!\!  \begin{bmatrix}
    n_0 \\
    n_1
  \end{bmatrix}\! ,
\end{align*}
where $c_0 \sim \mathcal{CN}\left(0, \frac{1}{N} \left\{\!\mathbb{E} \!\left[ |X|^2 \right] \!+\! \frac{2\rho_r}{1-\rho_r} \left( \!1 \!-\! \frac{1-\rho_r^{N}}{N (1-\rho_r)} \!\right) \!|\mathbb{E} \left[ X \right]\!|^2 \!\right\} \! \sigma_h^2 \! \right)$, $n_0 \sim \mathcal{CN}(0, \frac{1}{N} \sigma_n^2 )$ and $n_1 \sim \mathcal{CN}(0, \frac{1}{N} \sigma_n^2 )$. Taking cross-correlation between the first and second elements of $\mathbf{Z}^k_0$, we get:
\begin{align}
& \mathbb{E}\left[\sum\limits_{n=1}^{N} \frac{ y^k_{0,m}[n] }{N} 
    \sum\limits_{n=1}^{N} \frac{ (y^k_{0,m+1})^*[n] }{N}\right]  = \mathbb{E} \left[\left|c_0\right|^2\right] e^{-j\phi_1} + \mathbb{E} \left[c_0 n_1^*\right] + \mathbb{E} \left[c_0^* e^{-j\phi_1}  n_0\right] + \mathbb{E} \left[n_0 n_1^*\right]\nonumber\\
&= \frac{1}{N} \left\{\mathbb{E} \left[ |X|^2 \right] + \frac{2\rho_r}{1-\rho_r} \left( 1 - \frac{1-\rho_r^{N}}{N(1-\rho_r)} \right) |\mathbb{E} \left[ X \right]|^2 \right\}  \sigma_h^2 e^{-j\phi_1} \nonumber\\
&\implies e^{-j\phi_1} = \dfrac{\mathbb{E}\left[\sum\limits_{n=1}^{N} \frac{ y^k_{0,m}[n] }{N} 
    \sum\limits_{n=1}^{N} \frac{ (y^k_{0,m+1})^*[n] }{N}\right] }{\frac{1}{N} \left\{\mathbb{E} \left[ |X|^2 \right] + \frac{2\rho_r}{1-\rho_r} \left( 1 - \frac{1-\rho_r^{N}}{N (1-\rho_r)} \right) |\mathbb{E} \left[ X \right]|^2 \right\}  \sigma_h^2}.
\end{align}
A better estimate can be obtained by averaging over all the possible values of $m$ as follows:
\begin{align}
e^{-j\phi_1} =  \dfrac{\sum\limits_{m=0}^{M_r-2}\mathbb{E}\left[\sum\limits_{n=1}^{N} \frac{ y^k_{0,m}[n] }{N } 
    \sum\limits_{n=1}^{N} \frac{ (y^k_{0,m+1})^*[n] }{N}\right] }{\frac{ M_r-1}{N } \left\{\mathbb{E} \left[ |X|^2 \right] + \frac{2\rho_r}{1-\rho_r} \left( 1 - \frac{1-\rho_r^{N}}{N (1-\rho_r)} \right) |\mathbb{E} \left[ X \right]|^2 \right\}  \sigma_h^2}.
\end{align}
Since this averaging operation over  different antenna elements has independent noise terms, the accuracy of the estimate improves with the increasing value of $M_r$. The root mean square error (RMSE) of the DL AoA as a function of the SNR is shown in Fig. \ref{fig:rmse_AoA_DL}. As expected, the RMSE improves with the increasing SNR. The BER performance of the MA receiver over the RMSE values of interest is plotted in Fig. \ref{fig:BER_FShift_FS_Mr_AoA_Error_nocoding_SNR_N_2000}. Using a similar method for estimating the AoA of the BL does not result in good RMSE performance, which is mainly attributed to the interference from the DL. Hence,  alternate techniques are necessary to accurately estimate the AoA of the BL, and one potential method is to utilize the residual signal from the DL cancellation operation for the AoA estimation. Due to space limitations, it was not possible to include it in this paper and is hence left as a promising future work.

\begin{figure}
    \centering
    \begin{subfigure}[b]{0.45\textwidth}
        \centering
        \includegraphics [width=\linewidth]{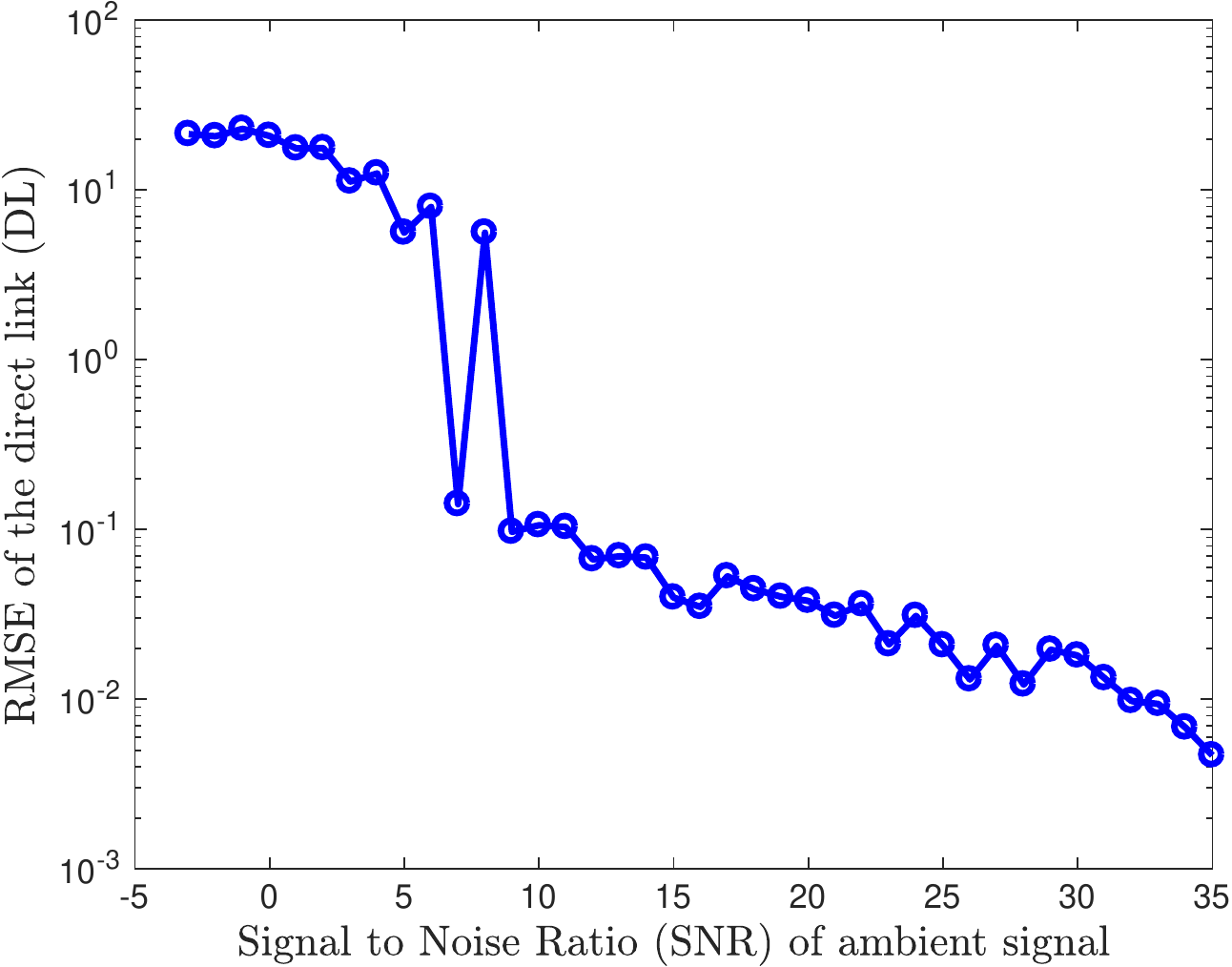}
	\caption{}\label{fig:rmse_AoA_DL}
    \end{subfigure}
    ~ 
    \begin{subfigure}[b]{0.45\textwidth}
        \centering
        \includegraphics [width=\linewidth]{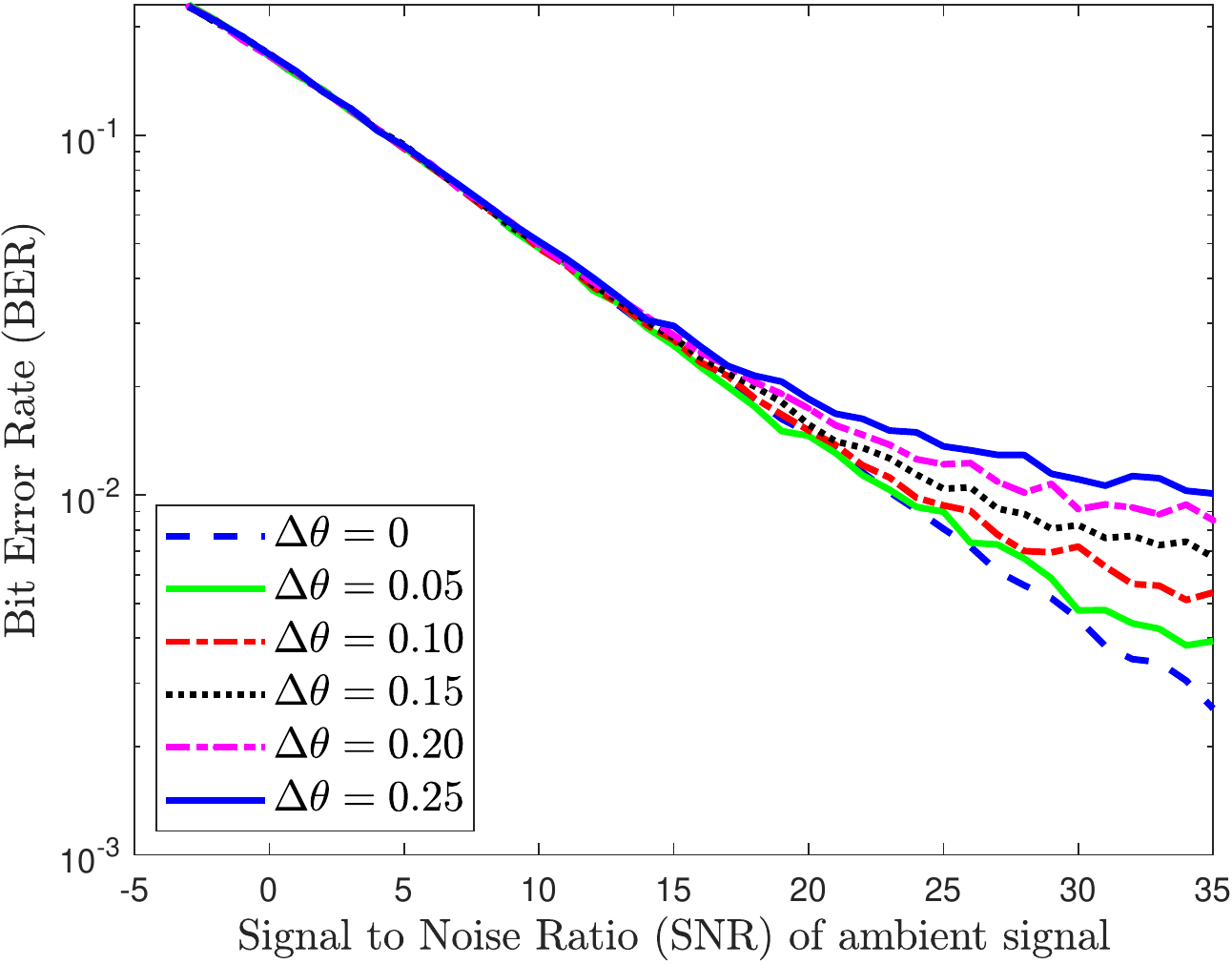}
         \caption{}\label{fig:BER_FShift_FS_Mr_AoA_Error_nocoding_SNR_N_2000}
    \end{subfigure}
    \caption{ (a) Root mean square error (RMSE) values of the estimated AoAs for the direct link (DL), and (b) BER performance comparison with estimation errors in AoA with $\Delta \theta =  \{ 0, 0.05, 0.10, 0.15,0.20, 0.25\}$.}
\end{figure}

}
\section{Numerical Results and Discussion} \label{sec:NumResults}

\begin{figure}
    \centering
    \begin{subfigure}[b]{0.45\textwidth}
        \centering
        \includegraphics [width=\linewidth]{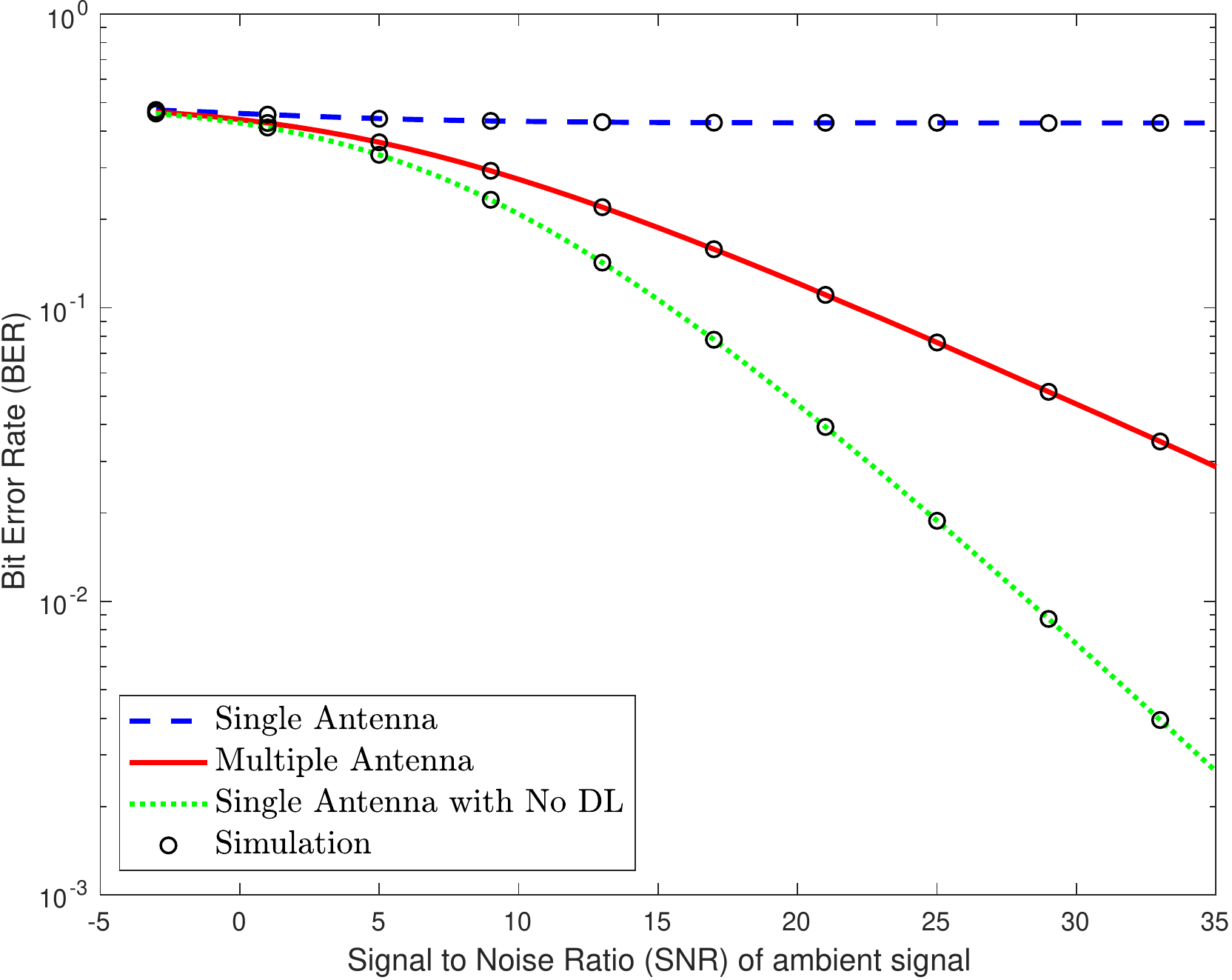}
         \caption{}\label{fig:BERSOMO}
    \end{subfigure}
    ~ 
    \begin{subfigure}[b]{0.46\textwidth}
        \centering
        \includegraphics [width=\linewidth]{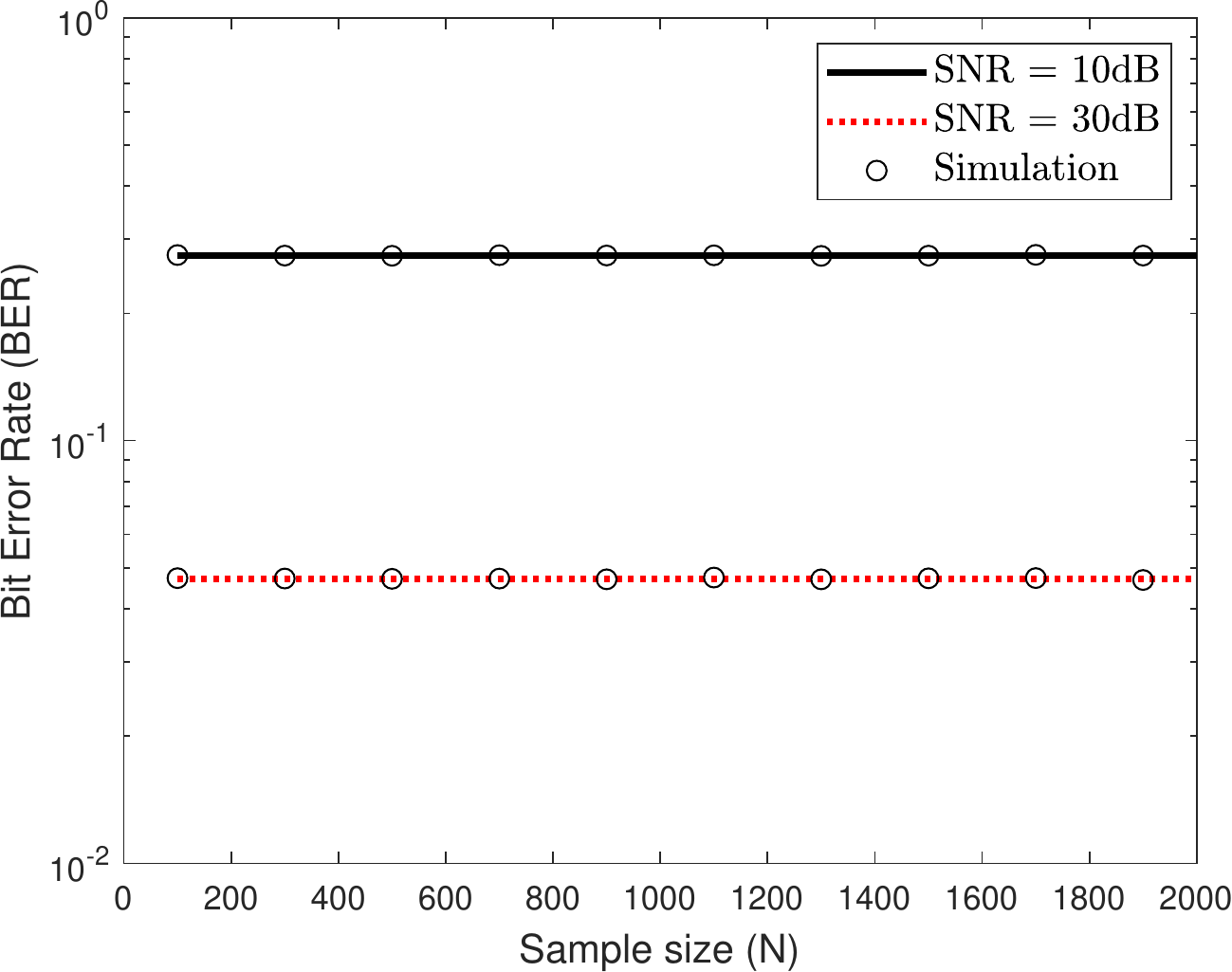}
	\caption{}\label{fig:BERSOMO_N}
    \end{subfigure}
    \caption{(a) BER comparison of the SA Rx and the MA Rx with $M_r=2$ under independent fading and/or ambient sequence with $\mathbb{E} \left[ X \right] = 0$, and comparison of the MA Rx with DL and the SA Rx without DL is also shown. (b) BER performance of the MA Rx with increasing $N$.}
\end{figure}

 \begin{figure}
    \centering
    \begin{subfigure}[b]{0.45\textwidth}
        \centering
        \includegraphics [width=\linewidth]{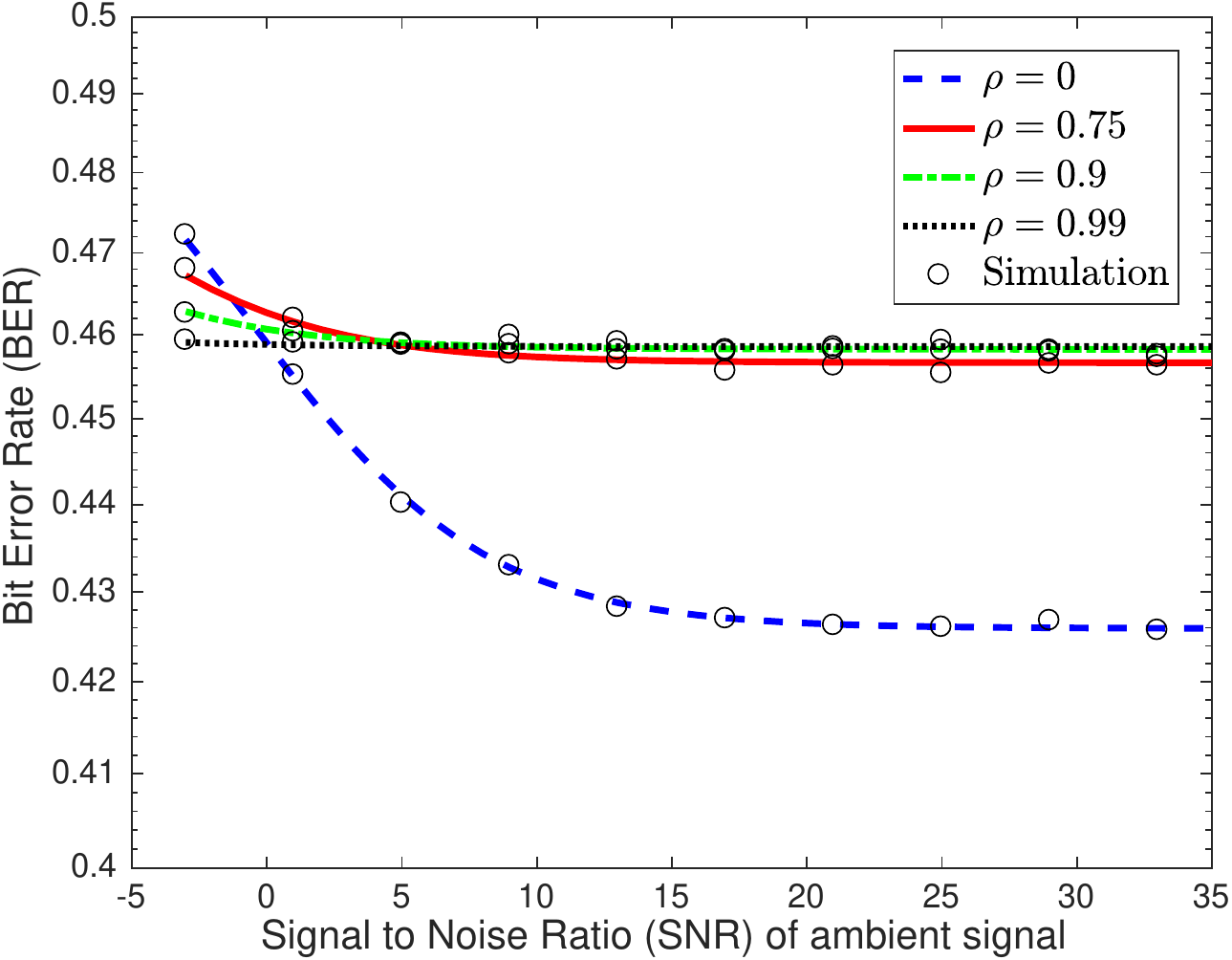}
	\caption{}\label{fig:BER_FS_rho_SA_variation_N5000}
    \end{subfigure}
    ~ 
    \begin{subfigure}[b]{0.45\textwidth}
        \centering
        \includegraphics [width=\linewidth]{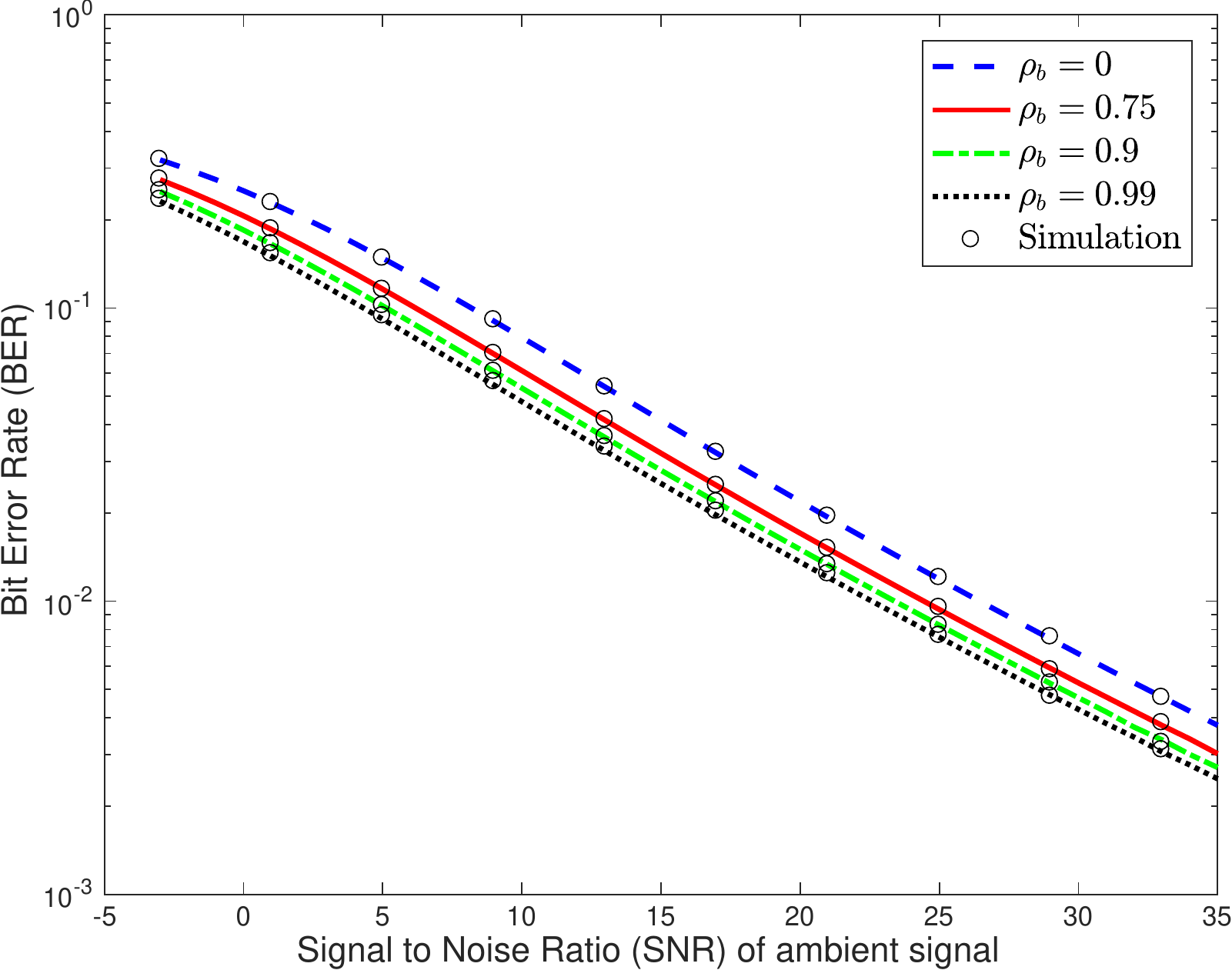}
         \caption{}\label{fig:BER_FShift_FS_Mr_rhovariation_nocoding_SNR_N_2000}
    \end{subfigure}
    \caption{ (a) BER performance and the error floor of the SA Rx for different $\rho$, (b) BER comparison of the MA Rx with $M_r=9$ for changing $\rho_b$, and the other parameters configured to $\rho_r = 0.5$ and $\rho_t = \rho_r \rho_b$.}
\end{figure}

In this section, the accuracy of  our analysis is verified by comparing with Monte-Carlo simulations. In addition, some useful system design insights are also provided. The reflection coefficient $\Gamma_1$ is configured appropriately to set the parameter $\alpha$ that will result in a signal attenuation of 1.1 dB, and the variance  of the fading gain $\sigma_{h}^2$ is set to $1$. The BER performance of the two receivers related to the special cases of independent fading $(\rho=0)$ and/or ambient sequence with zero expectation $(\mathbb{E} \left[ X \right] = 0)$, are compared in Fig. \ref{fig:BERSOMO}. We observe that with increasing SNR, the BER saturates quickly for a SA receiver without any further improvement. This behavior can be attributed to the dependence of a non-coherent detector on differences in the conditional variances of the received symbol. 
With the strong interference from power source, the variances of the two hypotheses scale similarly with increasing SNR. On the other hand, as shown in Fig. \ref{fig:BERSOMO}, the MA receiver can drastically improve the BER by removing the direct path from the ambient power source. In this case, BER decreases continuously without reaching any error floor. When the interference from the DL is removed in the MA receiver, only the variance of alternate hypothesis scales proportionally to the increasing SNR which ultimately results in the improved BER. Further, the average BER under these two cases is independent of the signal sample size $N$ as shown in Fig. \ref{fig:BERSOMO_N}. The effectiveness of the proposed DL cancellation technique is verified by comparing the BER of MA and SA receivers with and without the DL interference, respectively. As shown in Fig. \ref{fig:BERSOMO}, performance of the SA receiver without DL is better compared to the MA receiver with DL. This is expected because the BER of the MA receiver is averaged over the joint distribution of AoAs $\theta_1$ and $\theta_2$, and the performance is limited when the AoAs are similar. 

\begin{figure}
    \centering
    \begin{subfigure}[b]{0.45\textwidth}
        \centering
        \includegraphics [width=\linewidth]{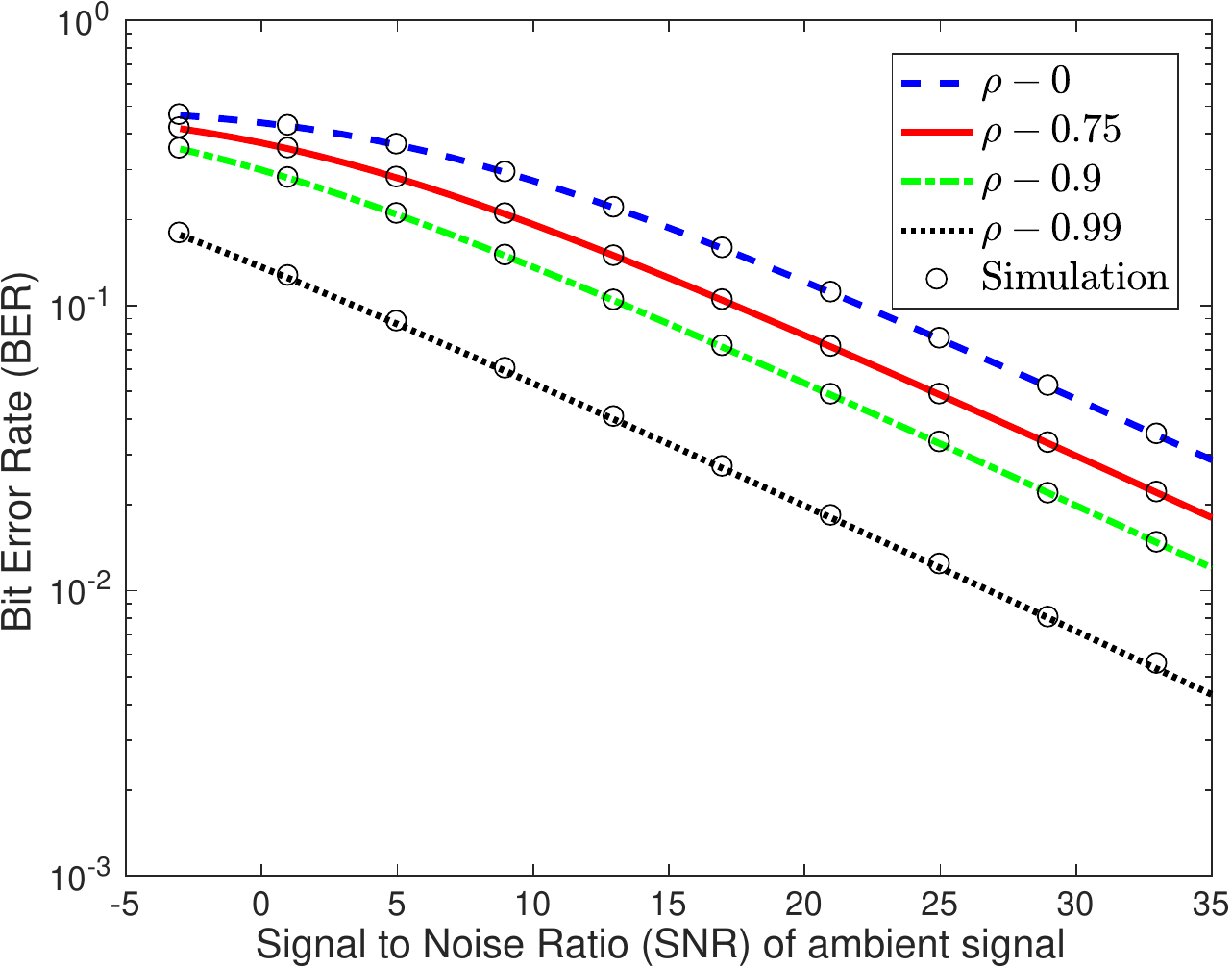}
         \caption{}\label{fig:BER_FS_rho_variation_N5000}
    \end{subfigure}
    ~ 
    \begin{subfigure}[b]{0.46\textwidth}
        \centering
        \includegraphics [width=\linewidth]{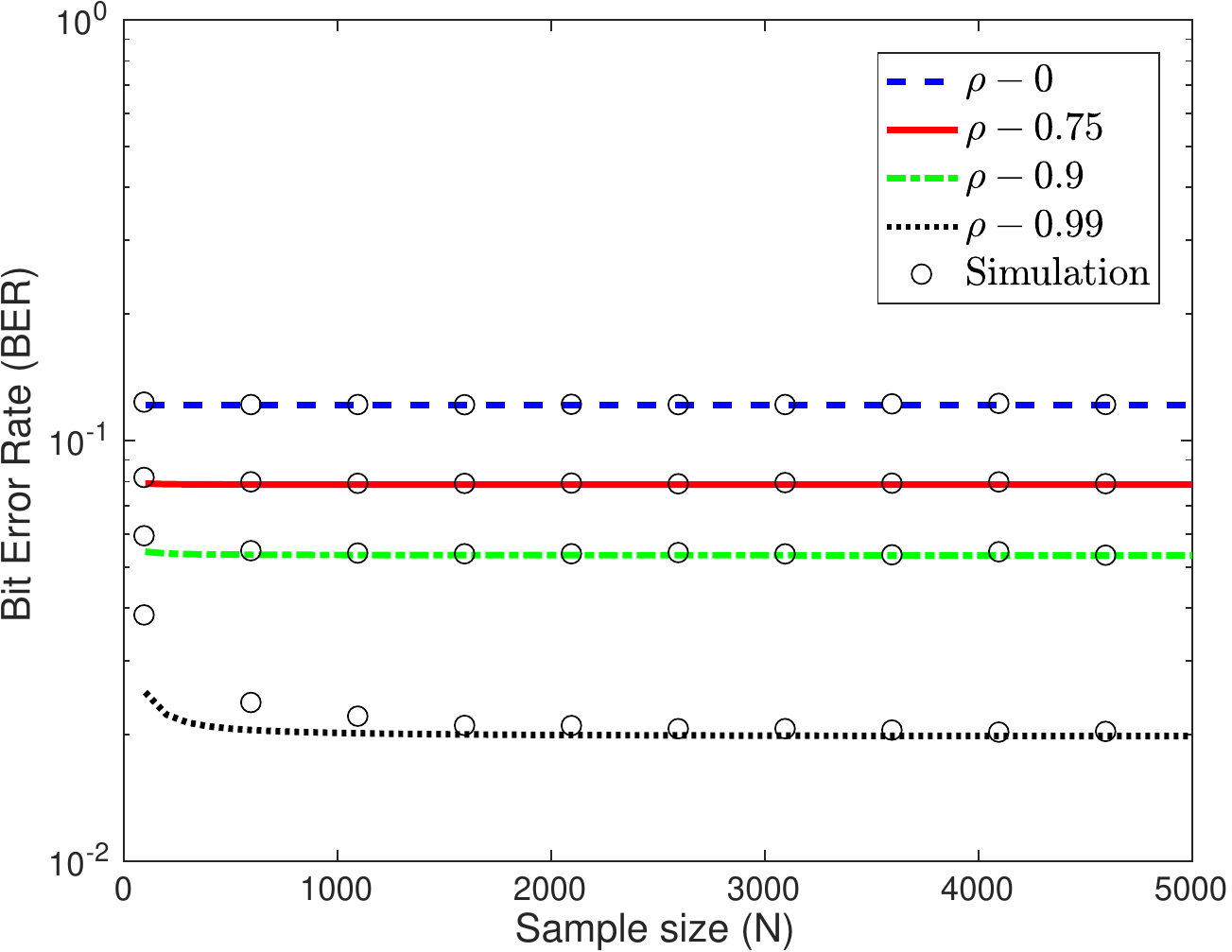}
	\caption{}\label{fig:BER_FS_rho_variation_N}
    \end{subfigure}
    \caption{ (a) BER vs SNR comparison of the MA Rx with $M_r=2$ for varying correlation factor $\rho$ and $N=5000$, (b) BER vs N comparison of the MA Rx with $M_r=2$ for changing correlation factor $\rho$ with $\rm SNR=20\, dB$.}
\end{figure}

\begin{figure}
    \centering
        \begin{subfigure}[b]{0.45\textwidth}
        \centering
        \includegraphics [width=\linewidth]{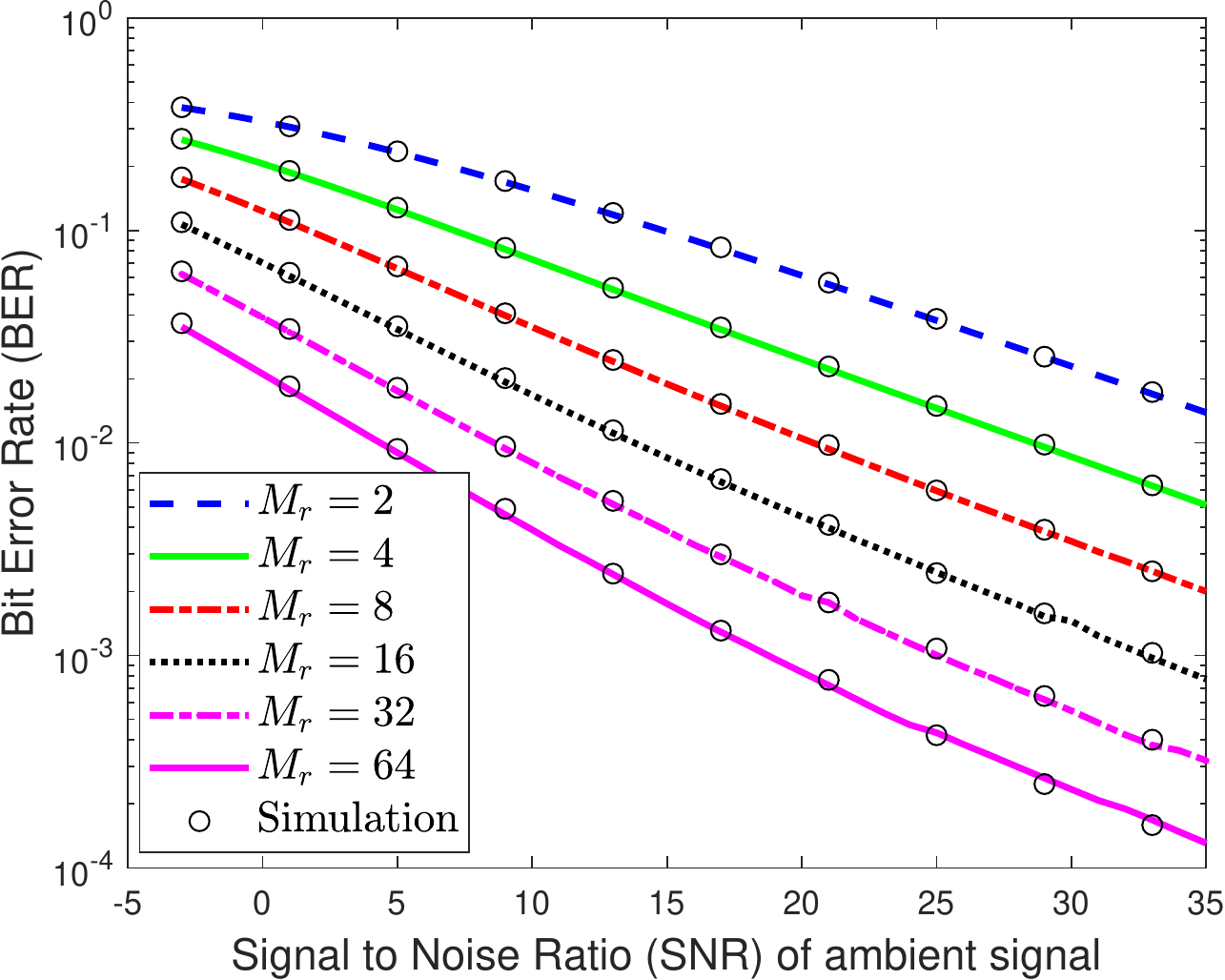}
         \caption{}\label{fig:BER_FShift_FS_Mr_variation_nocoding_SNR_N_2000}
    \end{subfigure}
    ~ 
\begin{subfigure}[b]{0.45\textwidth}
        \centering
        \includegraphics [width=\linewidth]{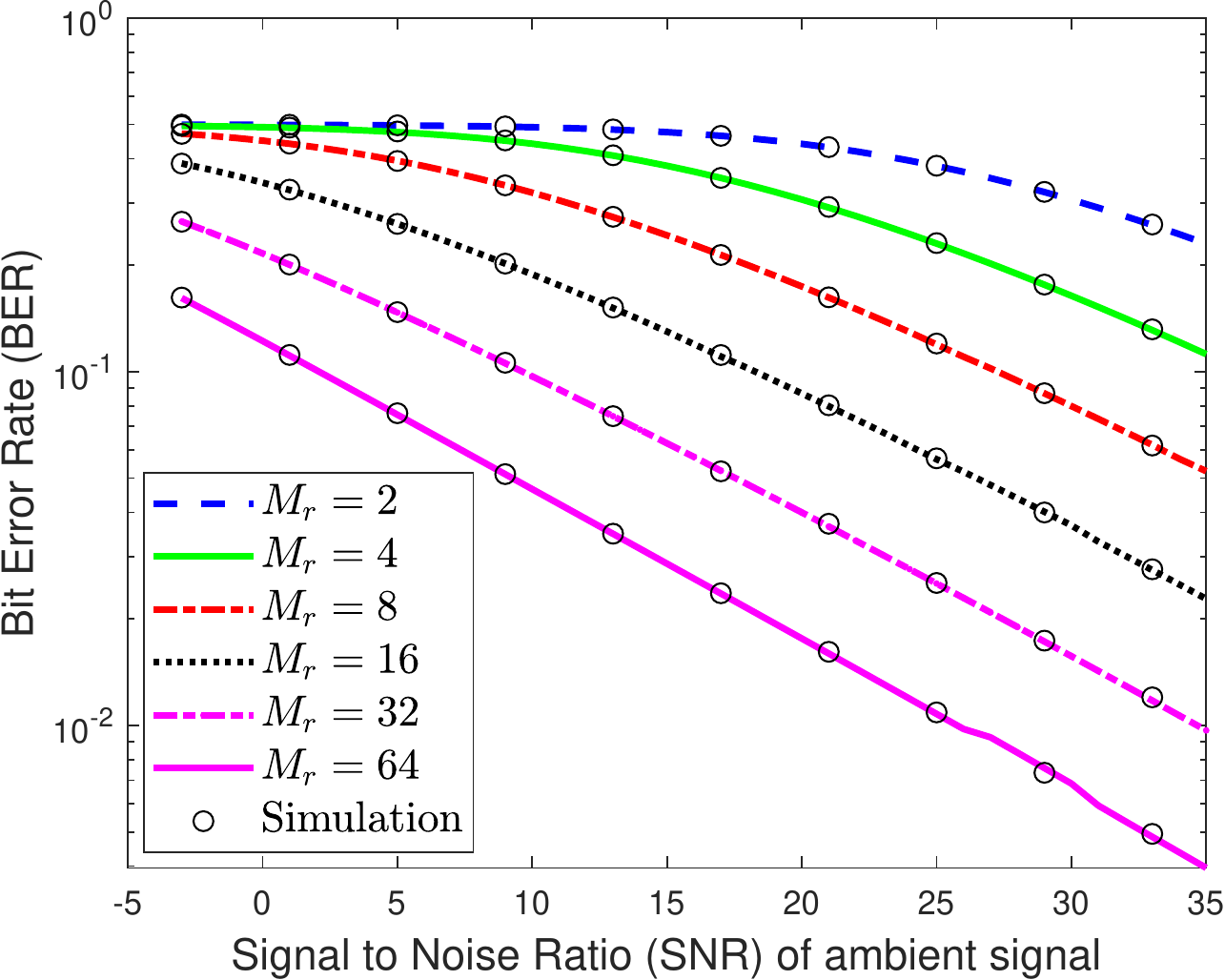}
         \caption{}\label{fig:BER_FShift_FS_Mr_variation_narrowAoA_nocoding_SNR_N_2000}
    \end{subfigure}
    \caption{ BER vs SNR comparison for changing antenna elements $M_r$ at the receiver with $\rho_r = 0.5, \rho_b=0.75, \rho_t = 0.38$ and $N=2000$: (a) uniformly distributed AoAs, and (b) narrowly distributed AoAs.}
\end{figure}

\begin{figure}
    \centering
    \begin{subfigure}[b]{0.45\textwidth}
        \centering
        \includegraphics [width=\linewidth]{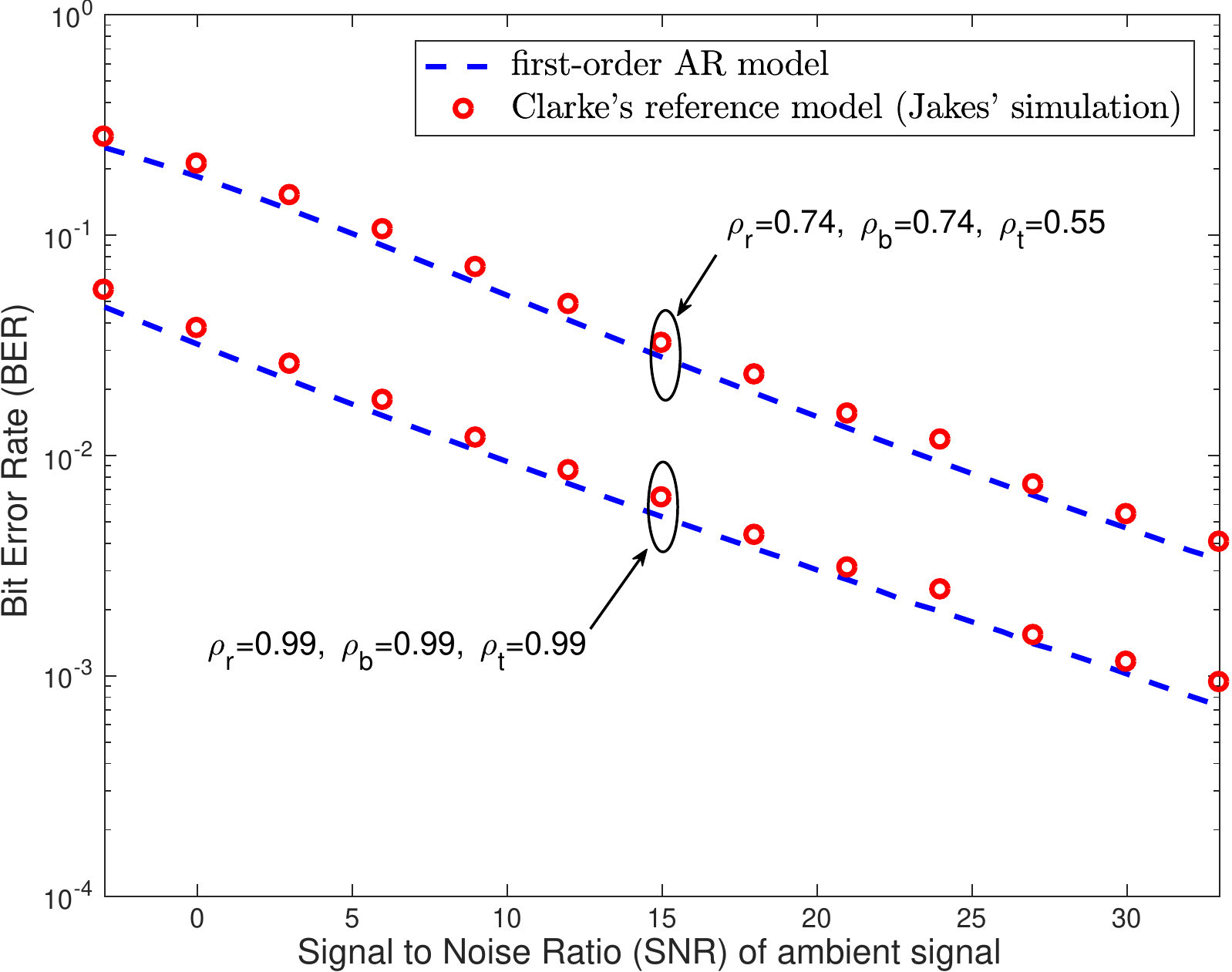}
         \caption{}\label{fig:BER_FShift_FS_Mr_rhovariation_jakes_nocoding_SNR_N_2000}
    \end{subfigure}
    ~ 
    \begin{subfigure}[b]{0.46\textwidth}
        \centering
        \includegraphics [width=\linewidth]{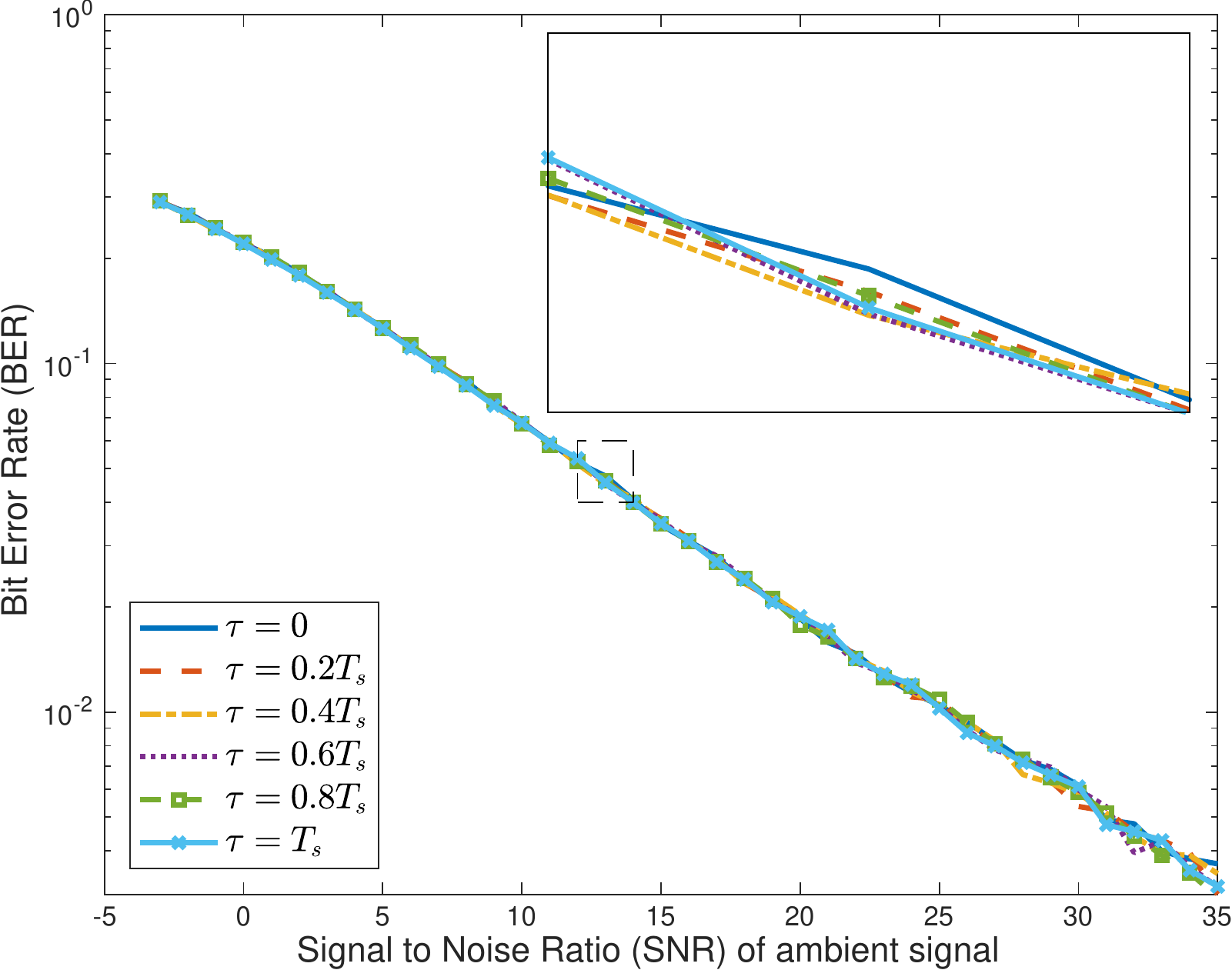}
	\caption{}\label{fig:BER_FShift_FS_Mr_rhovariation_delay_nocoding_SNR_N_2000}
    \end{subfigure}
    \caption{ (a) BER performance comparison of  AR model channel with that of the channel developed using Jakes' simulation model, and (b) Impact of changing timing error on the BER performance.}
\end{figure}

 The results for more general cases are discussed now. Unless specified explicitly for the particular plot, the values of different correlation factors $\rho_r, \rho_b$, and $\rho_t$ are all considered equal and represented as $\rho$. The error floor in a SA receiver decreases with correlation factor $\rho$, as shown in Fig. \ref{fig:BER_FS_rho_SA_variation_N5000}, and it can be inferred that a SA receiver is insufficient for non-coherent detection as the error floor values are very close to $0.5$, which corresponds to the BER of a naive hit/miss receiver. From Fig. \ref{fig:BER_FS_rho_SA_variation_N5000}, it can also be verified that the numerically obtained BER floor values of the SA receiver match with the asymptotic BER analytically derived in~\eqref{eq:asymBERM0SA}. The waterfall curve, as shown in Fig. \ref{fig:BER_FShift_FS_Mr_rhovariation_nocoding_SNR_N_2000},  validates our asymptotic BER analysis presented in~\eqref{eq:asymBERM0MA} for the MA receiver with unequal values for different correlation factors. 
The BER performance with increasing SNR in a MA receiver for different values of the correlation factor $\rho$ is presented in Fig. \ref{fig:BER_FS_rho_variation_N5000}, where it can be seen that the BER improves with increasing $\rho$. 
Likewise, the BER performance with increasing sample size $N$ for varying $\rho$ is shown in Fig. \ref{fig:BER_FS_rho_variation_N}, and interestingly the BER increases and saturates quickly with increasing $N$. However, as expected, there is an increasing mismatch between the simulated and theoretical results of BER at lower values of $N$ as the value of $\rho$ is increased. This mismatch occurs due to the need of a larger sample-size $N$ for the averaging operation, so that the simulation and theoretical results converge with increasing $\rho$. 
 The BER improvement observed with increasing $\rho$ and $N$ can be attributed to the increment in variance of the alternate hypothesis while the variance of the null hypothesis remains constant.  
  The antenna gain achieved with additional antennas is presented in Fig. \ref{fig:BER_FShift_FS_Mr_variation_nocoding_SNR_N_2000},  that shows around $8$ dB gain with the doubling of  antennas. The simulation result for the analysis in Remark \ref{rem:genmrspecial}, corresponding to the additional angular resolution achievable with antennas beyond two, is shown in Fig.  \ref{fig:BER_FShift_FS_Mr_variation_narrowAoA_nocoding_SNR_N_2000}. For this comparison, one can assume the AoA $\theta_1$ of the DL to be uniformly distributed between $(-\pi, \pi]$, and the AoA $\theta_2$ of the BL to be uniformly distributed with mean $\theta_1$ and width $\Delta \theta= 10^{\circ}$. 
 The results of the plot demonstrate that while the BER of the dual-antenna Rx is close to $0.5$, an antenna gain of around $9$ dB is achieved with the doubling of antennas in this case. The comparison between AR and Jakes' channel models discussed in Sec. \ref{sec:ChMod} is shown in Fig. \ref{fig:BER_FShift_FS_Mr_rhovariation_jakes_nocoding_SNR_N_2000}. Two scenarios are considered for comparison: 1) speed of PS and BTx are both $150$ kmph, and 2) speed of PS and BTx are both $5$ kmph. The corresponding values of the correlation factors for a signal of bandwidth $1.5$ KHz turns out to be: 1) $\rho_r= 0.74$, $\rho_b = 0.74$ and $\rho_t=0.55$, and 2) $\rho_r= 0.99$, $\rho_b = 0.99$ and $\rho_t=0.99$. The BER performance of our proposed approach for the AR model is similar to that of the Jakes' channel under these two scenarios. We checked many other scenarios and noticed a close match in all of them. We can therefore conclude that the simplified AR model approximates the actual complex time-selective channel very closely, while endowing tractability to the analysis. The approximation can be further improved by using a higher order AR process for modeling the time-selective fading channel.  
 Finally, as shown in Fig. \ref{fig:BER_FShift_FS_Mr_rhovariation_delay_nocoding_SNR_N_2000}, the impact of timing recovery errors is shown to be negligible, which corroborates our timing analysis in Section \ref{sec:SynchAmB}.
 
\section{Conclusion}

Ambient backscatter systems have mainly been studied for low mobility scenarios that are modeled using a block fading channel. While the block fading model is sufficient for stationary environments like home and office, a time-selective fading model is more suitable for non-stationary environments, such as roads and campuses. Therefore, in this paper, we have investigated the performance of an ambient backscatter system by studying the design and BER of a non-coherent detector under time-selective fading channels. 
To the best of our knowledge, this is the first work that has incorporated both non-coherent detection and time-selective fading into the ambient backscatter setup. Unlike the conventional architecture, which is implemented using the average of the energy of the received signal samples, a new receiver architecture based on the direct average of the signal samples is proposed. The new architecture is simpler to implement, robust to timing errors, and lends tractability to the asymptotic analysis. 
We have shown in the analysis that a BER floor exists for  the SA receiver due to the DL interference of the ambient power source, thereby resulting in an unacceptable performance. The BER is drastically improved using a MA receiver by tracking the AoA of the DL and using it to eliminate the interference. Further, having more than two receive antennas allows additional angular resolution, which can support applications where  the AoAs of the DL and BL links are very close. Though the BER in the time-selective fading improves with increasing signal sample-size, it saturates to an asymptotic value. 
Additionally, the BER is observed to improve with increasing temporal-correlation of the fading channel.  By comparing the BER, the simple first-order AR process is shown to be an effective approximation of the Clarke's reference model available for the time-selective fading channel. A natural extension of this work is to implement an ambient backscatter system that can function in a channel with multiple angular paths at the receiver. 

\appendix

\subsection{Proof of Lemma~\ref{lem:concineq}} \label{app:concineq}

The value of the summation $\sum\limits_{n_1 \neq n_2} \mspace{-8mu} \rho^{|n_1\mspace{-2mu} -n_2|}$, which is used in the subsequent steps is given by:
\begin{align}
 \sum\limits_{n_1 \neq n_2} \rho^{|n_1-n_2|} &= \frac{2\rho}{1-\rho} ( N - \frac{1-\rho^N}{1-\rho} ), & 0 \le \rho < 1. \label{eq:sumpn1n2}
\end{align}

The expectation of the sum sequence $S_N$ can be evaluated easily as follows:
\begin{align}
&\mathbb{E}\left[ S_N \right] \!=\! \mathbb{E}\left[\sum\limits_{n_1, n_2}  \rho^{|n_1-n_2|} \, x[n_1] x^*_m[n_2] \right] \!=\! \sum\limits_{n} \mathbb{E} \left[ |x[n]|^2 \right] \!+\!\! \sum\limits_{n_1 \neq n_2} \rho^{|n_1-n_2|} \,\mathbb{E} \left[ x[n_1] \right] \mathbb{E} \left[ x^*_m[n_2] \right] \nonumber\\
&\stackrel{(a)}{=} \sum\limits_{n}  \mathbb{E} \left[ |X|^2 \right] + \sum\limits_{n_1 \neq n_2} \rho^{|n_1-n_2|} \, |\mathbb{E} \left[ X \right]|^2 \stackrel{(b)}{=} N \mathbb{E} \left[ |X|^2 \right] + \frac{2\rho}{1-\rho} ( N \mspace{-3mu}-\mspace{-3mu} \frac{1-\rho^N}{1-\rho} ) |\mathbb{E} \left[ X \right]|^2,
\end{align}
where (a) and (b) follow from the assumption that the ambient sequence $x[n]$ is i.i.d., and the value of summation given in~\eqref{eq:sumpn1n2}, respectively. It can be easily observed that the expectation of this sum grows asymptotically of the order of $N$, meaning  $\mathbb{E}\left[ S_N \right] = \Theta[n]$. Using this, the expectation of $M_N = \frac{S_N}{N}$ can be shown to be a constant, whose value is given in~\eqref{eq:ExpMn}.

The variance of the sum sequence $S_N$ can first be simplified as given below:
\begin{align}
&{\rm Var}\left[ S_N \right] =  \mathbb{E} \left[ \left( \sum\limits_{i_1, j_1} \rho^{|i_1-j_1|} \, x[i_1] x^*_m[j_1] \right) \left( \sum\limits_{i_2, j_2} \rho^{|i_2-j_2|} \, x^*_m[i_2] x[j_2] \right) \right]  - \mathbb{E}\left[ S_N \right]^2 \nonumber\\
&= \mathbb{E} \Bigg[ \sum\limits_{i_1} \sum\limits_{i_2}   |x[i_1]|^2 |x[i_2]|^2 + 2 \sum\limits_{i_1} \sum\limits_{i_2 \neq j_2} \rho^{|i_2-j_2|} \, |x[i_1]|^2 x^*_m[i_2] x[j_2] \nonumber \\
& \quad \quad \quad + \sum\limits_{i_1 \neq j_1}  \sum\limits_{i_2 \neq j_2} \rho^{|i_1-j_1|+|i_2-j_2|} \, x[i_1] x^*_m[j_1] x^*_m[i_2] x[j_2]  \Bigg] - \mathbb{E}\left[ S_N \right]^2 \nonumber\\
&\stackrel{(c)}{=} \sum\limits_{i_1=i_2}  \mathbb{E} \left[|X|^4\right] +  \left\{ \sum\limits_{i_1 \neq i_2} 1 + \sum\limits_{i_1 \neq j_1} \rho^{2|i_1-j_1|} - ( \sum\limits_{i} 1 )^2 \right\} \left(\mathbb{E}\left[ |X|^2 \right]\right)^2 \nonumber\\
& + 2 \sum\limits_{i_1 \neq j_2} \rho^{|i_1-j_2|} \mathbb{E} \left[X (X^*)^2 \right] \mathbb{E} \left[X \right] + 2 \sum\limits_{i_1 \neq i_2} \rho^{|i_1-i_2|} \mathbb{E} \left[ (X)^2 X^*\right] \mathbb{E} \left[X^* \right] + \sum\limits_{i_1 \neq j_1} \rho^{2|i_1-j_1|}   \left|\mathbb{E} \left[ X^2 \right] \right|^2 \nonumber\\
&+ 2\left\{\sum\limits_{i_1 \neq i_2 \neq j_2} \rho^{|i_2-j_2|} +  \sum\limits_{i_1 \neq j_1\neq j_2} \rho^{|i_1-j_1|+|i_1-j_2|} - \sum\limits_{i, i_1 \neq i_2} \rho^{|i_1-i_2|} \right\} \mathbb{E} \left[ |X|^2 \right] \left|\mathbb{E} \left[X\right]\right|^2 \nonumber\\
&  +  \sum\limits_{i_1 \neq i_2 \neq j_1} \rho^{|i_1-j_1|+|i_2-j_1|} \mathbb{E} \left[ X^2 \right] \left(\mathbb{E} \left[X^*\right]\right)^2 + \sum\limits_{i_1 \neq j_1 \neq j_2} \rho^{|i_1-j_1|+|j_1-j_2|} \left(\mathbb{E} \left[ X^2 \right]\right)^* \left(\mathbb{E} \left[X\right]\right)^2 \nonumber\\
& + \left\{ \sum\limits_{i_1 \neq j_1 \neq i_2 \neq j_2} \rho^{|i_1-j_1|+|i_2-j_2|} - ( \sum\limits_{n_1 \neq n_2} \rho^{|n_1-n_2|} )^2 \right\}\left|\mathbb{E} \left[ X \right]\right|^4 \label{eq:varsnint},
\end{align}
where (c) follows from the piece-wise separation of different summations  by permuting the indices $i_1, i_2, j_1$ and $j_2$ of the first term, and the expansion of the second term $\mathbb{E}\left[ S_N \right]$.

The main objective here is to show that the variance also grows asymptotically of the order of $N$. The complete derivation of the variance expression is conceptually simple but tedious to present in a limited space. For this reason, we only provide a sketch of the proof, which is sufficient to understand the approach. 
Recall the assumption that the higher order moments of the sequences $x[n]$ upto the highest order present in~\eqref{eq:varsnint} are finite. With this assumption, it is sufficient to prove that the coefficient of each moment increases of the order of $N$.  The coefficient of $\mathbb{E} \left[|X|^4\right]$ is straightforward  to obtain and is given by $N$. Using~\eqref{eq:sumpn1n2}, it is again straightforward to show that $\sum\limits_{i_1 \neq j_1} \rho^{2|i_1-j_1|} $ is a function $N$, and the summations $ \sum\limits_{i_1 \neq i_2} 1 $ and $( \sum\limits_{i} 1 )^2$ are respectively given by $N^2$ and $N(N\mspace{-3mu}-\mspace{-3mu}1)$. Hence, the coefficient of $\left(\mathbb{E}\left[ |X|^2 \right]\right)^2$ is proportional to $N$ and increases asymptotically of the order of $N$. Then, the coefficients of $\mathbb{E} \left[X (X^*)^2 \right] \mathbb{E} \left[X \right], \mathbb{E} \left[ (X)^2 X^*\right] \mathbb{E} \left[X^* \right] $ and $  \left|\mathbb{E} \left[ X^2 \right] \right|^2$, given by either $\sum\limits_{i_1 \neq j_1} \rho^{|i_1-j_1|}$ or $\sum\limits_{i_1 \neq j_1} \rho^{2|i_1-j_1|}$, are already shown to be proportional to $N$. Similarly, the summation $\sum\limits_{i_1 \neq j_1\neq j_2} \mspace{-6mu}\rho^{|i_1-j_1|+|i_1-j_2|}$ can be evaluated by piece-wise categorization into different subsets and be shown to grow of the order of $N$. In addition, the summations $\sum\limits_{i_1 \neq i_2 \neq j_2} \mspace{-12mu}\rho^{|i_2-j_2|}$ and $\sum\limits_{i, i_1 \neq i_2}\mspace{-6mu} \rho^{|i_1-i_2|}$ can both be shown to have the same factor for $N^2$, and hence the coefficients of $\left|\mathbb{E} \left[ X^2 \right] \right|^2, \mathbb{E} \left[ X^2 \right] \left(\mathbb{E} \left[X^*\right]\right)^2$ and $\left(\mathbb{E} \left[ X^2 \right]\right)^* \left(\mathbb{E} \left[X\right]\right)^2$ all increase at the order of $N$. Finally, it can also be shown that $\sum\limits_{i_1 \neq j_1 \neq i_2 \neq j_2}\mspace{-12mu} \rho^{|i_1-j_1|+|i_2-j_2|}$ and $( \sum\limits_{n_1 \neq n_2} \mspace{-6mu}\rho^{|n_1-n_2|} )^2$ are both proportional to $N(N\mspace{-3mu}-\mspace{-3mu}1)$ with the same factor, which also means that $\left|\mathbb{E} \left[ X \right]\right|^4$ grows of the order of $N$. From this, we can conclude that 
 ${\rm Var}\left[ S_N \right] =\Theta
(N)$. As a consequence, the variance of $M_N = \frac{S_N}{N}$ will be decreasing at the rate of $1/N$ asymptotically. This completes the proof.

\subsection{Proof of Lemma~\ref{lem:CondPDFH0M1}} \label{app:CondPDFH0M1}

When conditioned on $x[n]$, a sample of the received signal under $\mathcal{H}_0$, as given in~\eqref{eq:nullhypeq}, is a complex Gaussian RV. As a result, the mean of the received samples can also be characterized as a complex Gaussian, \emph{albeit} the samples correlated with one another. Since the complex Gaussian RV is completely defined by its mean and  variance, we are just required to derive them. First, the conditional expectation and variance of an individual sample $y[n]$ can be derived as:
\begin{align*}
\mathbb{E} \left[y[n]\right] &= \mathbb{E} \left[\left( \rho_r^{n-1} h_r[1] + \sqrt{1-\rho_r^2} \left\{\sum\limits_{k=1}^{n-1}\rho_r^{n-k-1}g_r[k] \right\} \right) x[n] + w[n] \right] \\
&= \left( \rho_r^{n-1} \mathbb{E} \left[h_r[1]\right] + \sqrt{1-\rho_r^2} \left\{\sum\limits_{k=1}^{n-1}\rho_r^{n-k-1}\mathbb{E} \left[g_r[k]\right] \right\} \right) x[n] + \mathbb{E} \left[w[n]\right]= 0,\\
{\rm Var}\left[ y[n]\right] &= {\rm Var}\left[\rho_r^{n-1} h_r[1] x[n]\right] + \sum\limits_{k=1}^{n-1} {\rm Var}\left[ \sqrt{1-\rho_r^2} \left\{\rho_r^{n-k-1}g_r[k] x[n]\right\}\right]  + {\rm Var}\left[ w[n]\right]  \\
&= ( \rho_r^{2n-2} + \sum\limits_{k=1}^{n-1} (1-\rho_r^2) \rho^{2n-2k-2} ) \sigma_h^2 |x[n]|^2 +  \sigma_n^2 = \sigma_h^2 |x[n]|^2 +  \sigma_n^2.
\end{align*}
Similarly, the conditional covariance of any two distinct samples $y[i]$ and $y[j]$ is given by:
\begin{align*}
&{\rm Cov}\left[y[i], y[j] \right] \stackrel{(a)}{=}  \mathbb{E} \left[y[i] y^*[j] \right] \\
&= \mathbb{E} \Bigg[ (\rho_r^{i+j-2} |h_r[1]|^2  + \sqrt{1-\rho_r^2} \sum\limits_{k_2=1}^{j-1}\rho_r^{i+j-k_2-2}h_r[1]g_r^*[k_2] ) x[i] x^*[j]+ \rho_r^{i-1} h_r[1] x[i] w^*[j] \nonumber\\
& + ( \sqrt{1-\rho_r^2} \sum\limits_{k_1=1}^{i-1}\rho_r^{i+j-k_1-2}h_r^*[1]g_r[k_1]  + (1-\rho_r^2) \sum\limits_{k_1=1}^{i-1} \sum\limits_{k_2=1}^{j-1}\rho_r^{i+j-k_1-k_2-2}g_r[k_1]g_r^*[k_2] ) x[i] x^*[j] \nonumber\\
& + \sqrt{1-\rho_r^2} \sum\limits_{k_1=1}^{i-1}\rho_r^{i-k_1-1}w^*[j]g_r[k_1] x[i] + \rho_r^{j-1} h_r[1] x^*[j] w[i]  \nonumber\\
& + \sqrt{1-\rho_r^2} \sum\limits_{k_2=1}^{j-1}\rho_r^{j-k_2-1}w[i]g_r^*[k_2] x^*[j] + w[i] w^*[j]\Bigg]\\
&=  \sigma_h^2 (\rho_r^{i+j-2} + (1-\rho_r^2) \sum\limits_{k=1}^{\min(i,j)-1} \rho_r^{i+j-2k-2} ) x[i] x^*[j] = \sigma_h^2 \rho_r^{|j-i|} x[i] x^*[j],
\end{align*}
where (a) follows from zero valued conditional expectation of the signal samples. Using the above derivations, the conditional expectation and variance of $Z$ can be evaluated as follows:
\begin{align}
&\mathbb{E} \left[ Z \right] = \mathbb{E} \left[\frac{1}{N}\sum\limits_{n=1}^{N}y[n] \right]  = \frac{1}{N^2} \left( \mathbb{E} \left[\sum\limits_{n=1}^{N}y[n] \right]\right) = \frac{1}{N^2} \left( \sum\limits_{n=1}^{N}\mathbb{E} \left[y[n]\right]\right) = 0,\nonumber\\
&{\rm Var}\left[Z\right] = {\rm Var}\left[\frac{1}{N}\sum\limits_{n=1}^{N} y[n]\right] 
= \frac{1}{N^2} \left(\sum\limits_{n=1}^{N} {\rm Var}\left[y[n]\right] + \sum\limits_{n_1\neq n_2} {\rm Cov}\left[y[n_1], y[n_2] \right] \right) \nonumber\\
&= \frac{1}{N^2} (\sigma_h^2 \sum\limits_{n=1}^{N}|x[n]|^2  + N \sigma_n^2 + \sigma_h^2\sum\limits_{n_1\neq n_2} \rho_r^{|n_1-n_2|} \, x[n_1] x^*[n_2] ) = \frac{1}{N} \left(\sigma_h^2 {M_N} + \sigma_n^2 \right) \nonumber\\
&\stackrel{(b)}{\approx }\frac{1}{N} \left(\sigma_h^2 \mathbb{E}\left[ M_N \right] + \sigma_n^2 \right) = \frac{1}{N} (\sigma_h^2 \mathbb{E} \left[ |X|^2 \right] + \frac{2\rho_r}{1-\rho_r} ( 1 - \frac{1-\rho_r^N}{N(1-\rho_r)} ) |\mathbb{E} \left[ X \right]|^2 + \sigma_n^2 ), \label{eq:yvar}
\end{align}
where (b) results from the approximation of $M_N$ by its expectation, given in Lemma \ref{lem:concineq}.


\subsection{Proof of Lemma~\ref{lem:CondPDFH1M1}} \label{app:CondPDFH1M1}

Observe that when conditioned on the ambient signal $x[n]$, the three signal components of the received signal under the alternate hypothesis $\mathcal{H}_1$: (i)  direct signal from ambient source, (ii) backscatter signal, and (iii) receiver noise, are independent of each other.
\begin{align}
y[n] &=  h_r[n] x[n] + \underbrace{\alpha h_b[n] h_t[n] x[n]}_{y_b[n]} + w[n]
\end{align} 

This means that the expectation and variance of the sum can be derived using just the expectation and variance of each component. Since, we have already computed the expectation and variance of the direct signal and the receiver noise combination (in Lemma \ref{lem:CondPDFH1M1} for $\mathcal{H}_0$), it is now enough to compute the expectation and variance of the backscatter component $y_b[n]$.

To derive that, we further condition the signal on $h_b[n]$ since it will preserve and allow us to use the additive property of the Gaussian RVs. The conditional expectation and variance of an individual sample of the backscatter signal $y_b[n]$ and the conditional covariance of any two distinct samples $y[i]$ and $y[j]$ can be evaluated as:
\begin{align*}
\mathbb{E} \left[y_b[n] \right] &= \mathbb{E} \left[\alpha  h_{b}[n] h_t[n]  x[n] \right] = \alpha h_{b}[n]x[n]  \mathbb{E} \left[ h_t[n] \right] = 0,\\
{\rm Var}\left[ y_b[n]\right] &= {\rm Var}\left[ \alpha  h_{b}[n] h_t[n]  x[n]\right]  =  |\alpha|^2  |h_{b}[n]x[n]|^2 {\rm Var} \left[ h_t[n] \right] = |\alpha|^2 \sigma_h^2 |h_{b}[n]x[n]|^2 ,\\
{\rm Cov}\left[y_b[i], y_b[j] \right] 
&= |\alpha|^2  h_{b}[i] h_{b}^*[j] x[i] x^*[j] {\rm Cov}\left[h_t[i], h_t[j] \right]
=|\alpha|^2  \sigma_h^2 \rho_t^{|j-i|} h_{b}[i] h_{b}^*[j] x[i] x^*[j].
\end{align*}

The conditional expectation and variance of the mean of signal samples $y_b[n]$ can be determined from their corresponding expectation and variance of the individual samples as follows:
\begin{align}
&\mathbb{E} \left[\frac{1}{N}\sum\limits_{n=1}^{N}y_b[n] \right]  = \frac{1}{N} \left( \mathbb{E} \left[\sum\limits_{n=1}^{N}y_b[n] \right]\right) = \frac{1}{N} \left( \sum\limits_{n=1}^{N}\mathbb{E} \left[y_b[n]\right]\right) = 0, \nonumber\\
&{\rm Var}\left[\frac{1}{N}\sum\limits_{n=1}^{N} y_b[n]\right] 
= \frac{1}{N^2} \left(\sum\limits_{n=1}^{N} {\rm Var}\left[y_b[n]\right] + \sum\limits_{n_1\neq n_2} {\rm Cov}\left[y_b[n_1], y_b[n_2] \right] \right) \nonumber\\
&= \frac{1}{N^2} (|\alpha|^2 \sigma_h^2 \sum\limits_{n=1}^{N} |h_{b}[n]x[n]|^2  + |\alpha|^2 \sigma_h^2\sum\limits_{n_1\neq n_2} \rho_t^{|n_1-n_2|} \, h_{b}[n_1] h_{b}^*[n_2] x[n_1] x^*[n_2] ) \nonumber\\
&= \frac{1}{N} |\alpha|^2 \sigma_h^2 \underbrace{\frac{1}{N}\sum\limits_{1 \le n_1, n_2 \leq N}  \rho_t^{|n_1-n_2|} \, h_{b}[n_1] h_{b}^*[n_2] x[n_1] x^*[n_2] }_{M^b_N}. \label{eq:sumvarH1b}
\end{align}
The sequence $M^b_N$, similar to $M_N$, is a function of the sum variable of the ambient sequence $x[n]$ and can be shown to asymptotically converge to its expectation. This expected value of $M^b_N$ can be evaluated as follows:
\begin{align*}
&\mathbb{E}\left[ M^b_N \right] = \mathbb{E}\left[\frac{1}{N}\sum\limits_{1 \le n_1, n_2 \leq N}  \rho_t^{|n_1-n_2|} \, h_{b}[n_1] h_{b}^*[n_2] x[n_1] x^*[n_2]  \right] \nonumber\\
&= \frac{1}{N} \mathbb{E}\left[\sum\limits_{1 \le n \leq N}  |h_{b}[n]x[n]|^2 + \sum\limits_{n_1 \neq n_2} \rho_t^{|n_1-n_2|} \, h_{b}[n_1] h_{b}^*[n_2] x[n_1] x^*[n_2]\right]\nonumber\\
&= \frac{1}{N} ( \sum\limits_{1 \le n \leq N} \mathbb{E} \left[ |h_b[n]|^2 \right] \mathbb{E} \left[ |x[n]|^2 \right]+ \sum\limits_{n_1 \neq n_2} \rho_t^{|n_1-n_2|} \,  \mathbb{E} \left[ h_{b}[n_1] h_{b}^*[n_2] \right] \mathbb{E} \left[ x[n_1] \right]]\mathbb{E} \left[ x^*[n_1] \right] )\nonumber\\
 &\stackrel{(b)}{=}\! \sigma_h^2\!\!\!\sum\limits_{1 \le n \leq N}\!\!\! \frac{\mathbb{E} \!\left[ |X|^2 \right]}{N} \!+\! \sigma_h^2\!\! \sum\limits_{n_1 \neq n_2}\!\!\! (\rho_t \rho_b)^{|n_1-n_2|} \frac{|\mathbb{E} [X]|^2}{N} \!\!\stackrel{(c)}{=}\! \sigma_h^2 \mathbb{E} \left[ |X|^2 \right] \!+\! \sigma_h^2  \frac{2\rho_t \rho_b}{1\!\!-\!\!\rho_t \rho_b}\! \left(\!\! 1 \!-\! \frac{1\!-\! \rho_t^N \!\rho_b^{N}}{N(1\!\!-\!\!\rho_t \rho_b)}\! \right)\! \!|\mathbb{E} [X]|^2,
\end{align*}
where (b) follows from the assumption that the ambient sequence $x[n]$ is i.i.d. and the expectation of $h_{b}[n_1] h_{b}^*[n_2]$ which is given by $\sigma_h^2 \rho^{|n_1-n_2|}$, and (c) follows from the value of summation $\sum\limits_{n_1 \neq n_2} \rho^{2|n_1-n_2|}$ that can be derived using~\eqref{eq:sumpn1n2} in Lemma \ref{lem:concineq}.

The conditional variance of the mean of $y_b[n]$ can thus be approximated using $\mathbb{E}\left[ M^b_N \right]$ as:
\begin{align}
{\rm Var}\left[\frac{1}{N}\sum\limits_{n=1}^{N} y_b[n]\right] &\approx  \frac{1}{N} \left(|\alpha|^2 \sigma_h^2 \mathbb{E}\left[ M^b_N \right] \right)\nonumber\\
&= \frac{1}{N} \left(|\alpha|^2 \sigma_h^4 \mathbb{E} \left[ |X|^2 \right] + |\alpha|^2 \sigma_h^4  \frac{2\rho_t \rho_b}{1-\rho_t \rho_b} \!\left(\! 1 \!-\! \frac{1\!-\!\rho_t^N \! \rho_b^{N}}{N(1\!-\!\rho_t \rho_b)} \! \right) |\mathbb{E} \left[ X \right]|^2 \!\right) \label{eq:ybvar}. 
\end{align}
The final step is to obtain the variance of mean $Z$ of the signal samples under $\mathcal{H}_1$ by adding the individual variances in~\eqref{eq:yvar} and~\eqref{eq:ybvar} respectively. This completes the proof. 

%

\subsection{Proof of Theorem~\ref{thm:BERExpSOM1}} \label{app:BERExpSOM1}

The optimal decision rule for the receiver is evaluated through the comparison of the conditional PDFs of the null and alternate hypotheses $\mathcal{H}_0$ and $\mathcal{H}_1$ derived in Lemmas~\ref{lem:CondPDFH0M1} and~\ref{lem:CondPDFH1M1}, which is given by \cite{devineni19noncoherent}:
\begin{align*}
	\ln\left[f_{Z | \mathcal{H}_0}(z)\right] &\gtrless_{1}^{0} \ln\left[f_{Z | \mathcal{H}_1}(z)\right] \nonumber\\
	- \ln \left( {\rm Var}^{\rm SA}_0\right)-\dfrac{|z|^2} {{\rm Var}^{\rm SA}_0}  \gtrless_{1}^{0} -\ln \left({\rm Var}^{\rm SA}_1 \right)-\dfrac{|z|^2} {{\rm Var}^{\rm SA}_1} &\implies  |z|^2   \gtrless_{0}^{1}  \ln\left( \frac{{\rm Var}^{\rm SA}_1}{{\rm Var}^{\rm SA}_0}\right)\frac{{\rm Var}^{\rm SA}_1{\rm Var}^{\rm SA}_0}{{\rm Var}^{\rm SA}_1-{\rm Var}^{\rm SA}_0}, \label{eq: m1dr}
\end{align*}
where $z$ is the mean of signal samples. The value of the optimal detection threshold $T_{\rm SA}$ is given by the decision rule.

The decision rule of the optimal detection is only dependent on $|Z|^2$. The variable $|Z|^2$ is an exponential distributed RV, whose mean parameter equals the variance of the complex Gaussian. Assuming that the prior probabilities of the two hypotheses are equal, the conditional BER can be derived as:
\begin{align}
P_{\rm SA}(e) &= P(\mathcal{H}_0) P_{\rm SA}(e| \mathcal{H}_0) + P(\mathcal{H}_1) P_{\rm SA}(e| \mathcal{H}_1) \nonumber\\
&= \frac{1}{2} \left( Pr\left\{|Z|^2 > T_{\rm SA} | \mathcal{H}_0\right\} + Pr\left\{|Z|^2 < T_{\rm SA}| \mathcal{H}_1\right\} \right) \nonumber\\
&= \frac{1}{2} \left(1-F_{\rm exp}\left(T_{\rm SA}, {\rm Var}^{\rm SA}_0 \right) + F_{\rm exp}\left(T_{\rm SA}, {\rm Var}^{\rm SA}_1 \right) \right)
= \frac{1}{2} - \frac{1}{2} e^{{-\frac{T_{\rm SA}}{{\rm Var}^{\rm SA}_1}}} + \frac{1}{2}  e^{{-\frac{T_{\rm SA}}{{\rm Var}^{\rm SA}_0}}},\nonumber
\end{align}
where $F_{\rm Exp}(x,\lambda)$ is the cumulative distribution function of the exponential RV $|Z|^2$.

\subsection{Proof of Lemma~\ref{lem:SNRgain}} \label{app:SNRgain}
The antenna gain  $\tilde{\mathbf{a}}^* \mathbf{\hat{K}_{\tilde{W}}^{-1}} \tilde{\mathbf{a}}$ of the receiver is dependent on the inverse of $ \mathbf{\hat{K}_{\tilde{W}}}$, for which closed-form expression can be obtained. The matrix $\mathbf{\hat{K}_{\tilde{W}}}$ can be re-written as $\mathbf{\hat{K}_{\tilde{W}}}  = \mathbf{I}_{M_r -1} + \mathbf{J}_{M_r -1}$,
where $\mathbf{I}_{M_r -1} $ is an identity matrix and $\mathbf{J}_{M_r -1}$ is an all-ones matrix whose rank will be one. Therefore, $\mathbf{J}_{M_r -1}$ can be simplified using singular value decomposition (SVD) as $ \mathbf{u}_1 \sigma_1 \mathbf{v}_1^T$, where the unitary matrices  are given by $  \mathbf{u}_1
 =  \mathbf{v}_1 = \frac{-1}{\sqrt{M_r -1}} \begin{bmatrix}
   1 & 1 & \ldots & 1
  \end{bmatrix}^T$, and the non-zero singular value $\sigma_1 = M_r-1$. Due to the symmetry, this can be re-written in the form $\mathbf{J}_{M_r -1} = \mathbf{u}\mathbf{u}^T$, where   $\mathbf{u} = \begin{bmatrix}
   1 & 1 & \ldots & 1
  \end{bmatrix}^T$. Now, according to the Sherman-Morrison formula \cite{hager1989updating}, inverse of the sum of a invertible matrix $\mathbf{A}$ and the outer product $\mathbf{u}\mathbf{v}^T$ is given by $\left( \mathbf{A}+ \mathbf{u}\mathbf{v}^T \right)^{-1} = \mathbf{A}^{-1} - \dfrac{\mathbf{A}^{-1} \mathbf{u}\mathbf{v}^T \mathbf{A}^{-1}}{1+\mathbf{v}^T \mathbf{A}^{-1} \mathbf{u}}$. The Sherman-Morrison formula is considered as a special case of the Woodbury matrix identity \cite{hager1989updating}. Using this, the inverse of $\mathbf{\hat{K}_{\tilde{W}}}$ can be derived as:
\begin{align}
\mathbf{\hat{K}_{\tilde{W}}^{-1}} &
= \mathbf{I}_{M_r -1} - \frac{\mathbf{u}\mathbf{u}^T}{1+ \mathbf{u}^T\mathbf{u}} 
= \mathbf{I}_{M_r -1} - \frac{\mathbf{J}_{M_r -1}}{M_r} .
\end{align}
The expression of the SNR gain $\tilde{\mathbf{a}}^* \mathbf{\hat{K}_{\tilde{W}}^{-1}} \tilde{\mathbf{a}}$ can be simplified as follows:
\begin{align*}
& \tilde{\mathbf{a}}^* \mathbf{\hat{K}_{\tilde{W}}^{-1}} \tilde{\mathbf{a}} = \begin{bmatrix}
     e^{-j(\phi_2-\phi_1)}-1\\
    \vdots\\
    e^{-j(M_r-1)(\phi_2-\phi_1)}-1
  \end{bmatrix}^T  \begin{bmatrix}
   \frac{M_r-1}{M_r} & \frac{-1}{M_r} & \ldots & \frac{-1}{M_r} \\
   \vdots & \vdots & \ddots & \vdots\\
   \frac{-1}{M_r}& \frac{-1}{M_r} & \ldots & \frac{M_r-1}{M_r}
  \end{bmatrix} \begin{bmatrix}
     e^{j(\phi_2-\phi_1)}-1\\
    \vdots\\
    e^{j(M_r-1)(\phi_2-\phi_1)}-1
  \end{bmatrix}\\
 &= \sum\limits_{i=1}^{M_r-1} \left[ e^{j i(\phi_2-\phi_1)}-1\right] \left[ e^{-j i(\phi_2-\phi_1)}-1\right] - \frac{S_{M_r-1} S_{M_r-1}^*}{M_r}  = - S_{M_r-1} - S_{M_r-1}^* - \frac{S_{M_r-1} S_{M_r-1}^*}{M_r} ,
\end{align*} 
where $S_{M_r-1}  \!= \!\sum\limits_{i=1}^{M_r-1} \left[ e^{j i(\phi_2-\phi_1)}-1\right]$ is the summation of all the elements in the weight vector. Since, $S_{M_r-1} $ is a geometric sum it can be simplified, and the sum $S_{M_r-1} + S_{M_r-1}^*$ and product $S_{M_r-1} S_{M_r-1}^*$ can be derived as following:
\begin{align*}
S_{M_r-1} + S_{M_r-1}^* &= 2 \frac{\sin  \left((M_r-1)\frac{\phi_2-\phi_1}{2} \right) }{\sin \left(\frac{\phi_2-\phi_1}{2}  \right)} \cos \left( \frac{M_r}{2} (\phi_2-\phi_1)\right) - 2(M_r-1)\\
S_{M_r-1} S_{M_r-1}^* &= \frac{\sin^2  \left((M_r-1)\frac{\phi_2-\phi_1}{2} \right) }{\sin^2 \left(\frac{\phi_2-\phi_1}{2} \right)} + (M_r-1)^2 \nonumber\\
 & \quad \quad \quad - 2(M_r-1) \frac{\sin  \left((M_r-1)\frac{\phi_2-\phi_1}{2} \right) }{\sin \left(\frac{\phi_2-\phi_1}{2}  \right)}  \cos \left( \frac{M_r}{2} (\phi_2-\phi_1)\right).
\end{align*}
Using these simplifications, the final expression for the SNR gain can be determined as follows:
\begin{align*}
\tilde{\mathbf{a}}^* \mathbf{\hat{K}_{\tilde{W}}^{-1}} \tilde{\mathbf{a}} 
&= M_r\!-\!\frac{1}{M_r} \!- \!\frac{2}{M_r} \frac{\sin  \left((M_r\!-\!1)\frac{\phi_2-\phi_1}{2} \right) }{\sin \left(\frac{\phi_2-\phi_1}{2}  \right)}  \cos \!\left(\! \frac{M_r}{2} (\phi_2-\phi_1)\!\right) \!  - \!\frac{1}{M_r} \frac{\sin^2  \left((M_r\!-\!1)\frac{\phi_2-\phi_1}{2} \right) }{\sin^2 \left(\frac{\phi_2-\phi_1}{2} \right)}.
\end{align*}
\subsection{Proof of Lemma~\ref{lem:CondPDFMA}} \label{app:CondPDFMA}
The effective signal $y_{\rm eff}[n]$, given in~\eqref{eq:effsignalMr}, under $\mathcal{H}_0$ is a complex Gaussian RV with variance $ \sigma_n^2$. Hence, the mean $Z$ of the received samples under $\mathcal{H}_0$ is a complex Gaussian RV with variance ${\rm Var}^{\rm MA}_0 = \frac{ \sigma_n^2}{N}$. On the other hand, $y_{\rm eff}[n]$ under $\mathcal{H}_1$ is the sum of a scaled version of the backscatter signal $y_{b}[n]$ in Lemma \ref{lem:CondPDFH1M1} with the same receiver noise variance. Using the procedure similar to the ones in Lemmas \ref{lem:CondPDFH0M1} and \ref{lem:CondPDFH1M1}, the mean $Z$ of the received samples under $\mathcal{H}_1$ can also be shown to follow a complex Gaussian distribution, the variance of which is given by $${\rm Var}^{\rm MA}_1 = \frac{ G |\alpha|^2 \sigma_h^4 \left\{ \mathbb{E} \left[ |X|^2 \right] +  \frac{2\rho_t\rho_b}{1-\rho_t \rho_b} \left( 1 - \frac{1-\rho_t^{N}\rho_b^{N}}{N(1-\rho_t\rho_b)} \right) |\mathbb{E} \left[ X \right]|^2  \right\}+ \sigma_n^2}{N}.$$

\subsection{Proof of Theorem~\ref{thm:BERExpMO}} \label{app:BERExpMO}
By comparing the conditional PDFs of the two hypotheses given in~\eqref{eq:MOHi}, the optimal detection threshold $T_{\rm MA}$ can be obtained. The conditional BER, evaluated using a procedure similar to the one used in the case of SA receiver, is a function of the phase-offsets of the DL and BL links, and the average BER is obtained by marginalizing the conditional BER over the variables $\theta_1$ and $\theta_2$. The assumption here is that $\theta_1$ and $\theta_2$ are i.i.d. and uniformly distributed over $(-\pi, \pi]$, and the final expression in the result can be obtained by marginalizing over this range of $\theta_1$ and $\theta_2$. One can choose more complex distributions of AoAs to model different scenarios.  

%
%

{ \def\baselinestretch{1.3}
\bibliographystyle{IEEEtran}
	\bibliography{ambnoncoherent}
}

\end{document}

%% file: notation.tex
\def\nb0{{\mathbf{0}}}
\def\nb1{{\mathbf{1}}}

\newtheorem{lemma}{Lemma}

\newtheorem{theorem}{Theorem}

\newtheorem{remark}{Remark}










%% file: Draft_v2a.bbl
\begin{thebibliography}{10}
\providecommand{\url}[1]{#1}
\csname url@samestyle\endcsname
\providecommand{\newblock}{\relax}
\providecommand{\bibinfo}[2]{#2}
\providecommand{\BIBentrySTDinterwordspacing}{\spaceskip=0pt\relax}
\providecommand{\BIBentryALTinterwordstretchfactor}{4}
\providecommand{\BIBentryALTinterwordspacing}{\spaceskip=\fontdimen2\font plus
\BIBentryALTinterwordstretchfactor\fontdimen3\font minus
  \fontdimen4\font\relax}
\providecommand{\BIBforeignlanguage}[2]{{%
\expandafter\ifx\csname l@#1\endcsname\relax
\typeout{** WARNING: IEEEtran.bst: No hyphenation pattern has been}%
\typeout{** loaded for the language `#1'. Using the pattern for}%
\typeout{** the default language instead.}%
\else
\language=\csname l@#1\endcsname
\fi
#2}}
\providecommand{\BIBdecl}{\relax}
\BIBdecl

\bibitem{devineni20multi}
J.~K. Devineni and H.~S. Dhillon, ``Multi-antenna non-coherent detection of
  ambient backscatter under time-selective fading,'' \emph{Proc., IEEE
  Globecom}, Dec. 2020.

\bibitem{devineni19noncoherent}
------, ``Non-coherent signal detection and bit error rate for an ambient
  backscatter link under fast fading,'' \emph{Proc., IEEE Globecom}, Dec. 2019.

\bibitem{shyam13}
V.~Liu, A.~Parks, V.~Talla, S.~Gollakota, D.~Wetherall, and J.~R. Smith,
  ``Ambient backscatter: Wireless communication out of thin air,'' \emph{Proc.,
  ACM SIGCOMM}, Aug. 2013.

\bibitem{shyam16}
B.~Kellogg, V.~Talla, S.~Gollakota, and J.~R. Smith, ``Passive {W}i-{F}i:
  Bringing low power to {W}i-{F}i transmissions,'' \emph{Symposium on NSDI},
  vol.~16, pp. 151--164, Mar. 2016.

\bibitem{katti2015}
D.~Bharadia, K.~Joshi, M.~Kotaru, and S.~Katti, ``Back{F}i: high throughput
  {W}i-{F}i backscatter,'' \emph{Proc., ACM SIGCOMM}, pp. 283--296, Aug. 2015.

\bibitem{dhillon2015wide}
H.~S. Dhillon, H.~Huang, and H.~Viswanathan, ``Wide-area wireless communication
  challenges for the internet of things,'' \emph{IEEE Commun. Mag.}, vol.~55,
  no.~2, pp. 168--174, Feb. 2017.

\bibitem{devineni2019ambient}
J.~K. Devineni and H.~S. Dhillon, ``Ambient backscatter systems: Exact average
  bit error rate under fading channels,'' \emph{IEEE Trans. Green Commun. and
  Networking}, vol.~3, no.~1, pp. 11--25, Mar. 2019.

\bibitem{Wang15}
K.~Lu, G.~Wang, F.~Qu, and Z.~Zhong, ``Signal detection and {BER} analysis for
  {RF}-powered devices utilizing ambient backscatter,'' \emph{Proc., Intl.
  Conf. on Wireless Commun. \& Sig. Proc. (WCSP)}, Oct. 2015.

\bibitem{chintha15}
G.~Wang, F.~Gao, Z.~Dou, and C.~Tellambura, ``Uplink detection and {BER}
  analysis for ambient backscatter communication systems,'' \emph{Proc., IEEE
  Globecom}, Dec. 2015.

\bibitem{chintha16}
G.~Wang, F.~Gao, R.~Fan, and C.~Tellambura, ``Ambient backscatter communication
  systems: Detection and performance analysis,'' \emph{IEEE Trans. Commun.},
  vol.~64, no.~11, pp. 4836 -- 4846, Nov. 2016.

\bibitem{gao16}
J.~Qian, F.~Gao, G.~Wang, S.~Jin, and H.~Zhu, ``Semi-{C}oherent {D}etection and
  {P}erformance {A}nalysis for {A}mbient {B}ackscatter {S}ystem,'' \emph{IEEE
  Trans. Commun.}, vol.~65, no.~12, Dec. 2017.

\bibitem{gao17}
------, ``Noncoherent {D}etections for {A}mbient {B}ackscatter {S}ystem,''
  \emph{IEEE Trans. Wireless Commun.}, vol.~16, no.~3, Mar. 2017.

\bibitem{hu2015}
Y.~Liu, Z.~Zhong, G.~Wang, and D.~Hu, ``Uplink detection and {BER} performance
  for wireless communication systems with ambient backscatter and multiple
  receiving antennas,'' \emph{Proc., Intl. Conf. on Commun. and Networking in
  China (ChinaCom)}, pp. 79 -- 84, Aug. 2015.

\bibitem{chintha16vtc}
T.~Zeng, G.~Wang, Y.~Wang, Z.~Zhong, and C.~Tellambura, ``Statistical
  {C}ovariance {B}ased {S}ignal {D}etection for {A}mbient {B}ackscatter
  {C}ommunication {S}ystems,'' \emph{Proc., IEEE Veh. Technology Conf. (VTC)},
  Sep. 2016.

\bibitem{yang17}
G.~Yang, Y.-C. Liang, R.~Zhang, and Y.~Pei, ``Modulation in the {A}ir:
  {B}ackscatter {C}ommunication over {A}mbient {OFDM} {C}arrier,'' \emph{IEEE
  Trans. Commun.}, vol.~66, no.~3, Mar. 2018.

\bibitem{el2019noncoherent}
M.~A. El~Mossallamy, M.~Pan, R.~J{\"a}ntti, K.~G. Seddik, G.~Y. Li, and Z.~Han,
  ``Noncoherent backscatter communications over ambient {OFDM} signals,''
  \emph{IEEE Trans. Commun.}, 2019.

\bibitem{tao2018symbol}
Q.~Tao, C.~Zhong, H.~Lin, and Z.~Zhang, ``Symbol detection of ambient
  backscatter systems with manchester coding,'' \emph{IEEE Trans. Wireless
  Commun.}, vol.~17, no.~6, pp. 4028--4038, 2018.

\bibitem{wang17}
Y.~Liu, G.~Wang, Z.~Dou, and Z.~Zhong, ``New {C}oding and {D}etection {S}chemes
  for {A}mbient {B}ackscatter {C}ommunication {S}ystems,'' \emph{IEEE Access},
  Mar. 2017.

\bibitem{zhao2018blind}
W.~Zhao, G.~Wang, S.~Atapattu, and B.~Ai, ``Blind channel estimation in ambient
  backscatter communication systems with multiple-antenna reader,'' in
  \emph{2018 IEEE/CIC Intl. Conf. on Commun. in China (ICCC)}, 2018, pp.
  320--324.

\bibitem{guo2018noncoherent}
H.~Guo, Q.~Zhang, D.~Li, and Y.-C. Liang, ``Noncoherent multiantenna receivers
  for cognitive backscatter system with multiple {RF} sources,'' \emph{arXiv
  preprint, arXiv:1808.04316}, 2018.

\bibitem{zhang2018constellation}
Q.~Zhang, H.~Guo, Y.-C. Liang, and X.~Yuan, ``Constellation learning-based
  signal detection for ambient backscatter communication systems,'' \emph{IEEE
  J. Sel. Areas Commun.}, vol.~37, no.~2, pp. 452--463, 2018.

\bibitem{guo2018exploiting}
H.~Guo, Q.~Zhang, S.~Xiao, and Y.-C. Liang, ``Exploiting multiple antennas for
  cognitive ambient backscatter communication,'' \emph{IEEE Internet of Things
  Journal}, vol.~6, no.~1, pp. 765--775, 2018.

\bibitem{darsena2018joint}
D.~Darsena, G.~Gelli, and F.~Verde, ``Joint channel estimation, interference
  cancellation, and data detection for ambient backscatter communications,'' in
  \emph{2018 IEEE 19th Intl. Workshop on SPAWC}, 2018, pp. 1--5.

\bibitem{shyam14}
A.~N. Parks, A.~Liu, S.~Gollakota, and J.~R. Smith, ``Turbocharging ambient
  backscatter communication,'' \emph{Proc., ACM SIGCOMM}, pp. 1--12, Aug. 2014.

\bibitem{duan2019hybrid}
R.~Duan, E.~Menta, H.~Yi{\u{g}}itler, and R.~J{\"a}ntti, ``Hybrid beamformer
  design for high dynamic range ambient backscatter receivers,'' \emph{arXiv
  preprint, arXiv:1901.05323}, 2019.

\bibitem{varshney17}
A.~Varshney, O.~Harms, C.~Perez-Penichet, C.~Rohner, F.~Hermans, and T.~Voigt,
  ``Lorea: A backscatter architecture that achieves a long communication
  range,'' \emph{Proc., ACM on Embedded Network Sensor Systems (SenSys 17)},
  no.~50, Nov. 2017.

\bibitem{wang2017fm}
A.~Wang, V.~Iyer, V.~Talla, J.~R. Smith, and S.~Gollakota, ``{FM} backscatter:
  Enabling connected cities and smart fabrics,'' in \emph{Symposium on
  Networked Systems Design and Implementation (NSDI 17)}, 2017, pp. 243--258.

\bibitem{raleigh1994characterization}
G.~Raleigh, S.~N. Diggavi, A.~F. Naguib, and A.~Paulraj, ``Characterization of
  fast fading vector channels for multi-antenna communication systems,'' in
  \emph{Proc. of 1994 28th Asilomar Conf. on Signals, Systems and Computers},
  vol.~2.\hskip 1em plus 0.5em minus 0.4em\relax IEEE, 1994, pp. 853--857.

\bibitem{sayeed2002deconstructing}
A.~M. Sayeed, ``Deconstructing multiantenna fading channels,'' \emph{IEEE
  Trans. Signal Processing}, vol.~50, no.~10, pp. 2563--2579, 2002.

\bibitem{dhillon2015wireless}
H.~S. Dhillon and G.~Caire, ``Wireless backhaul networks: Capacity bound,
  scalability analysis and design guidelines,'' \emph{IEEE Trans. Wireless
  Commun.}, vol.~14, no.~11, pp. 6043--6056, 2015.

\bibitem{adhikary2013joint}
A.~Adhikary, J.~Nam, J.-Y. Ahn, and G.~Caire, ``Joint spatial division and
  multiplexing-the large-scale array regime,'' \emph{IEEE Trans. Information
  Theory}, vol.~59, no.~10, pp. 6441--6463, 2013.

\bibitem{adhikary2014joint}
A.~Adhikary, E.~Al~Safadi, M.~K. Samimi, R.~Wang, G.~Caire, T.~S. Rappaport,
  and A.~F. Molisch, ``Joint spatial division and multiplexing for mm-wave
  channels,'' \emph{IEEE J. Sel. Areas Commun.}, vol.~32, no.~6, pp.
  1239--1255, 2014.

\bibitem{adhikary2015massive}
A.~Adhikary, H.~S. Dhillon, and G.~Caire, ``Massive-mimo meets hetnet:
  Interference coordination through spatial blanking,'' \emph{IEEE J. Sel.
  Areas Commun.}, vol.~33, no.~6, pp. 1171--1186, 2015.

\bibitem{stuber2017principles}
G.~St{\"u}ber, \emph{Principles of Mobile Communication}.\hskip 1em plus 0.5em
  minus 0.4em\relax Springer International Publishing, 2017.

\bibitem{baddour2005autoregressive}
K.~E. Baddour and N.~C. Beaulieu, ``Autoregressive modeling for fading channel
  simulation,'' \emph{IEEE Trans. Wireless Commun.}, vol.~4, no.~4, pp.
  1650--1662, 2005.

\bibitem{liu2002space}
Z.~Liu, X.~Ma, and G.~B. Giannakis, ``Space-time coding and kalman filtering
  for time-selective fading channels,'' \emph{IEEE Trans. Commun.}, vol.~50,
  no.~2, pp. 183--186, 2002.

\bibitem{komninakis2002multi}
C.~Komninakis, C.~Fragouli, A.~H. Sayed, and R.~D. Wesel, ``Multi-input
  multi-output fading channel tracking and equalization using kalman
  estimation,'' \emph{IEEE Tran. on Signal Processing}, vol.~50, no.~5, pp.
  1065--1076, 2002.

\bibitem{ghandour2012use}
S.~Ghandour-Haidar, L.~Ros, and J.-M. Brossier, ``On the use of first-order
  autoregressive modeling for rayleigh flat fading channel estimation with
  kalman filter,'' \emph{Signal Processing}, vol.~92, no.~2, pp. 601--606,
  2012.

\bibitem{wang1996verifying}
H.~S. Wang and P.-C. Chang, ``On verifying the first-order markovian assumption
  for a rayleigh fading channel model,'' \emph{IEEE Trans. Vehicular
  Technology}, vol.~45, no.~2, pp. 353--357, 1996.

\bibitem{fuschini2008analytical}
F.~Fuschini, C.~Piersanti, F.~Paolazzi, and G.~Falciasecca, ``Analytical
  approach to the backscattering from {UHF RFID} transponder,'' \emph{IEEE
  Antennas and Wireless Propagation Letters}, vol.~7, pp. 33--35, 2008.

\bibitem{balanis1997antenna}
C.~A. Balanis, ``Antenna theory: analysis and design, john wiley \& sons,''
  \emph{New York}, 1997.

\bibitem{bletsas2010improving}
A.~Bletsas, A.~G. Dimitriou, and J.~N. Sahalos, ``Improving backscatter radio
  tag efficiency,'' \emph{IEEE Trans. Microwave Theory Tech}, vol.~58, no.~6,
  pp. 1502--1509, 2010.

\bibitem{darsena2019noncoherent}
D.~Darsena, ``Noncoherent detection for ambient backscatter communications over
  {OFDM} signals,'' \emph{IEEE Access}, vol.~7, pp. 159\,415--159\,425, 2019.

\bibitem{guruacharya2019optimal}
S.~Guruacharya, X.~Lu, and E.~Hossain, ``Optimal non-coherent detector for
  ambient backscatter communication system,'' \emph{arXiv preprint,
  arXiv:1911.10105}, 2019.

\bibitem{dtse}
D.~Tse and P.~Viswanath, ``Fundamentals of wireless communication,''
  \emph{Cambridge University Press}, 2005.

\bibitem{mengali2013synchronization}
U.~Mengali, \emph{Synchronization techniques for digital receivers}.\hskip 1em
  plus 0.5em minus 0.4em\relax Springer Science \& Business Media, 2013.

\bibitem{pahlavan2005wireless}
K.~Pahlavan and A.~H. Levesque, \emph{Wireless information networks}.\hskip 1em
  plus 0.5em minus 0.4em\relax John Wiley \& Sons, 2005.

\bibitem{hager1989updating}
W.~W. Hager, ``Updating the inverse of a matrix,'' \emph{SIAM review}, vol.~31,
  no.~2, pp. 221--239, 1989.

\end{thebibliography}
